%% file: 00_main.tex
\documentclass[a4paper,fleqn]{cas-dc}

\usepackage[numbers]{natbib}
\usepackage{amsmath,amssymb,amsfonts}
\usepackage{algorithmic}
\usepackage{graphicx}
\usepackage{textcomp}
\usepackage{fancyvrb}
\usepackage{subcaption}
\usepackage{adjustbox}
\usepackage{booktabs}
\usepackage{xspace}
\usepackage{todonotes}
\usepackage{cleveref}
\usepackage{ragged2e}
\usepackage{rotating}
\usepackage{multirow}
\usepackage{xstring}
\usepackage{mfirstuc}
\usepackage{longtable}
\usepackage[most]{tcolorbox}
\usepackage{caption}
\usepackage{xcolor}
\usepackage{listings}
\usepackage{fontawesome5}
\usepackage{soul}

\definecolor{rolecolor}{RGB}{0,102,204}   
\definecolor{taskcolor}{RGB}{0,153,51}    
\definecolor{formatcolor}{RGB}{204,102,0} 
\definecolor{infocolor}{RGB}{96, 96, 96}
\definecolor{fewshotcolor}{RGB}{138,43,226} 
\definecolor{labelbg}{rgb}{0.80, 0.90, 1.0}  
\definecolor{labeltext}{rgb}{0.0, 0.2, 0.5} 

\definecolor{pbg}{rgb}{0.88, 1.0, 0.88}
\definecolor{ptext}{rgb}{0.0, 0.3, 0.0}
\newcommand{\plabel}[2]{%
    \sethlcolor{pbg}%
    \textcolor{ptext}{\hl{#1 (\textbf{#2})}}{}
}

\definecolor{nbg}{rgb}{1.0, 0.88, 0.88}
\definecolor{ntext}{rgb}{0.5, 0.0, 0.0}
\newcommand{\nlabel}[2]{%
    \sethlcolor{nbg}%
    \textcolor{ntext}{\hl{#1 (\textbf{#2})}}{}%
}

\definecolor{mbg}{rgb}{1.0, 0.95, 0.80}
\definecolor{mtext}{rgb}{0.4, 0.2, 0.0}
\newcommand{\mlabel}[2]{%
    \sethlcolor{mbg}%
    \textcolor{mtext}{\hl{#1 (\textbf{#2})}}{}%
}

\definecolor{agreement}{HTML}{91cf60}

\DefineVerbatimEnvironment{PromptBlock}{Verbatim}{
    fontfamily=tt,
    fontsize=\small,
    xleftmargin=1em,
    frame=single,
    rulecolor=\color{black},
    framesep=0.5em
}

\lstdefinestyle{javastyle}{
  language=Java,
  basicstyle=\ttfamily\scriptsize,
  numbersep=8pt,
  breaklines=true,
  tabsize=2,
  showstringspaces=false,
  keywordstyle=\color{blue!70!black},
  commentstyle=\color{gray!70},
  stringstyle=\color{orange!70!black},
    frame=tb,
    rulecolor=\color{gray!50},
    framesep=5pt
}

\lstdefinestyle{python}{
  language=python,
  basicstyle=\ttfamily\footnotesize,
  numbersep=8pt,
  breaklines=true,
  tabsize=2,
  showstringspaces=false,
  keywordstyle=\color{blue!70!black},
  commentstyle=\color{gray!70},
  stringstyle=\color{orange!70!black},
    frame=tb,
    rulecolor=\color{gray!50},
    framesep=5pt
}

\newcommand{\promptsec}[1]{{\scriptsize\textbf{--#1--}}}

\newcommand{\kalpha}[0]{Krippendorff's~$\alpha$}

\newcommand{\qwCode}{\textsc{Qwen2.5-Coder 14B}\xspace}
\newcommand{\qwCodeShort}{\textsc{Qwen}\xspace}
\newcommand{\mistDev}{\textsc{Dev\-stral Small 1.1}\xspace}
\newcommand{\mistDevShort}{\textsc{Dev\-stral}\xspace}
\newcommand{\dskCode}{\textsc{Deep\-Seek-Coder-V2}\xspace}
\newcommand{\dskCodeShort}{\textsc{Deep\-Seek}\xspace}

\input{cmd/errors}

\newcounter{ctstep}                 
\newcommand{\ctTotal}{1}            

\definecolor{ctFrame}{HTML}{D0D7DE}
\definecolor{ctBack}{HTML}{FFFFFF}
\definecolor{ctHeader}{HTML}{1F2328}
\definecolor{ctHeaderFg}{HTML}{F6F8FA}
\definecolor{ctCodeBack}{HTML}{F6F8FA}
\definecolor{ctNote}{HTML}{FAFBFC}

\newtcolorbox{codetour@box}[1][]{%
  enhanced,
  colframe=ctFrame, colback=ctBack,
  boxrule=0.6pt, arc=3pt,
  left=4pt, right=4pt, top=4pt, bottom=4pt,
  before skip=4pt, after skip=8pt,
  pad at break*=4pt,
  #1
}

\newcommand{\pinktt}[1]{\textcolor{magenta}{\texttt{#1}}}

\newenvironment{codetour}[3]{%
  \begin{figure*}[]
  \renewcommand{\ctTotal}{#1}%
  \setcounter{ctstep}{0}%
  \def\ctCaption{#2}\def\ctLabel{#3}%
  \begin{codetour@box}
}{%
  \end{codetour@box}
  \caption{\ctCaption}\label{\ctLabel}
  \end{figure*}
}

\newtcolorbox{codestep@inner}[1]{%
  enhanced, breakable, 
  colframe=ctHeader, colback=ctCodeBack,
  coltitle=ctHeaderFg,
  fonttitle=\ttfamily\small\bfseries,
    title={\ttfamily#1},       
  boxrule=0pt, arc=2pt,
  left=2pt, right=2pt, top=4pt, bottom=4pt,
  before skip=4pt, after skip=4pt,
}

\newenvironment{codestep}[1]{%
  \stepcounter{ctstep}%
  \begin{codestep@inner}{#1}%
}{%
  \end{codestep@inner}%
}

\newtcolorbox{stepnote}{%
  enhanced, breakable,  
  colframe=ctFrame, colback=ctNote,
  boxrule=0.4pt, arc=2pt,
    left=4pt, right=4pt, top=2pt, bottom=2pt, 
  before skip=4pt, after skip=4pt,         
  fontupper=\scriptsize,                   
     before upper={\scriptsize\raggedright\noindent\textbf{Step \thectstep/\ctTotal}\\[2pt]\ignorespaces},
}

\definecolor{quotebg}{RGB}{250, 250, 250}
\definecolor{quotetext}{rgb}{0.15, 0.15, 0.15}  
\newcommand{\quotep}[1]{%
    \sethlcolor{quotebg}%
    \textcolor{quotetext}{\textit{\hl{``#1''}}}%
}

\newcommand{\commit}[3]{\href{https://github.com/#1/#2/commit/#3}{#3}}

\newtcolorbox{stepbox}{
    enhanced,
    colback=ctNote,             
    colframe=ctFrame,    
    boxrule=1pt,               
    arc=3pt,                   
    auto outer arc,
    left=6pt,                  
    right=6pt,                 
    top=6pt,                   
    bottom=6pt,                
    fontupper=\small,          
}

\definecolor{diffadd}{RGB}{0,128,0}
\definecolor{diffremove}{RGB}{200,0,0}
\definecolor{diffhunk}{RGB}{0,0,200}
\definecolor{difffile}{RGB}{128,0,128}
\definecolor{diffctx}{RGB}{80,80,80}
\definecolor{bgcolor}{RGB}{248,248,248}

\lstdefinestyle{diffstyle}{
    basicstyle=\ttfamily\scriptsize,
    backgroundcolor=\color{bgcolor},
    breaklines=true,
    frame=single,
    rulecolor=\color{black!30},
    numbers=none,
    showstringspaces=false,
    moredelim=[il][\color{difffile}\bfseries]{diff\ },
    moredelim=[il][\color{difffile}\bfseries]{index\ },
    moredelim=[il][\color{difffile}\bfseries]{---\ },
    moredelim=[il][\color{difffile}\bfseries]{+++\ },
    moredelim=[il][\color{diffhunk}\bfseries]{@@},
    moredelim=[il][\color{diffadd}]{+},
    moredelim=[il][\color{diffremove}]{-},
    moredelim=[il][\color{diffctx}]{\ }
}

\lstdefinestyle{trace}{
    basicstyle=\ttfamily\scriptsize,
    backgroundcolor=\color{bgcolor},
    breaklines=true,
    frame=single,
    rulecolor=\color{black!30},
    numbers=none
}

\def\tsc#1{\csdef{#1}{\textsc{\lowercase{#1}}\xspace}}
\tsc{WGM}
\tsc{QE}
\tsc{EP}
\tsc{PMS}
\tsc{BEC}
\tsc{DE}

\begin{document}
\let\WriteBookmarks\relax
\def\floatpagepagefraction{1}
\def\textpagefraction{.001}
\shorttitle{Human Factors in AI-generated Code Tours}
\shortauthors{Balfroid et~al.}

\title [mode = title]{How Developers Experience Debugging Unfamiliar Codebases with Code Tours Generated and Evaluated by Local LLMs}

\tnotetext[1]{This research was supported by the ARIAC project (No. 2010235), funded by the Service Public de Wallonie (SPW Recherche). We gratefully acknowledge the participants for their valuable contributions to this study.}

\author[]{Martin Balfroid}[orcid=0000-0002-1318-1184, bioid=1]
\cormark[1]
\ead{martin.balfroid@unamur.be}
\credit{Conceptualization, Methodology, Software, Validation, Formal analysis, Data Curation, Investigation, Visualization, Writing -- original draft, Writing -- review \& editing}

\author[]{Julien Albert}[orcid=0000-0001-6279-5601, bioid=2]
\credit{
Validation,
Formal analysis, 
Data Curation, 
Writing -- review \& editing
}

\author[]{Dzenatan Aliti}[bioid=3]
\credit{
Formal analysis, 
Data Curation 
}

\author[]{Xavier Devroey}[bioid=4]
\credit{
Supervision, Validation, Writing -- review \& editing, Project administration, Funding acquisition, Resources}

\author[]{Beno\^it Vanderose}[bioid=5]
\credit{
Supervision, Validation, Writing -- review \& editing, Project administration, Funding acquisition, Resources}

\cortext[cor1]{Corresponding author}


\newcommand{\abstractsection}[1]{%
\par\smallskip\noindent\textbf{#1:}
}

\newcounter{finding}
\renewcommand{\thefinding}{F\arabic{finding}}

\crefname{finding}{finding}{findings}
\Crefname{finding}{Finding}{Findings}

\newcommand{\finding}[2]{%
  \par\addvspace{\medskipamount}%
  \begin{tcolorbox}[
    enhanced,
    colback=gray!5,
    colframe=gray!5,     
    frame hidden,          
    boxrule=0pt,
    sharp corners,
    borderline north={0.5pt}{0pt}{black},
    borderline south={0.5pt}{0pt}{black},
    breakable=false,
    left=6pt, right=6pt, top=4pt, bottom=4pt,
  ]
    \refstepcounter{finding}%
    \noindent\textbf{\faSearch~[\thefinding] #1}\label{finding:#2}%
  \end{tcolorbox}%
  \addvspace{\bigskipamount}%
}

\newcounter{insight}
\renewcommand{\theinsight}{H\arabic{insight}}

\newcommand{\insight}[2]{%
  \par\addvspace{\medskipamount}%
  \begin{tcolorbox}[
    enhanced,
    colback=gray!5,
    colframe=gray!5,     
    frame hidden,          
    boxrule=0pt,
    sharp corners,
    borderline north={0.5pt}{0pt}{black},
    borderline south={0.5pt}{0pt}{black},
    breakable=false,
    left=6pt, right=6pt, top=4pt, bottom=4pt,
  ]
    \refstepcounter{insight}%
    \noindent\textbf{\faLightbulb~[\theinsight] #1}\label{insight:#2}%
  \end{tcolorbox}%
  \addvspace{\bigskipamount}%
}

\newcommand{\lightrule}[0]{
\arrayrulecolor{lightgray}\cmidrule(l){8-9}\arrayrulecolor{black}
}

\begin{abstract}
\normalfont 
\abstractsection{Context}
Code tours are interactive, onboarding documentation that guide developers through a codebase. Large Language Models (LLMs) can automatically synthesize code tours. Prior work on code tour generation has not examined developer experience or trust calibration when debugging unfamiliar codebases with code tours generated and evaluated by open-weight LLMs.
\abstractsection{Objectives}
This study surveys how the properties of components in open-weight LLM-authored code tours influence developers' experiences when debugging unfamiliar codebases.
\abstractsection{Method} 
We built a pipeline that generated and evaluated code tours from real reproducible bugs. Twenty-six developers with varying backgrounds participated in a user study. In total, 26 code tours were authored from real Java bugs mined from 2025 GitHub commits, with each tour independently judged by two different LLMs, resulting in 52 evaluated configurations. Participants thought aloud as they explored each tour. Three authors qualitatively coded the interviews to identify recurring themes.
\abstractsection{Results}
Developers generally preferred tours that scaled detail with the code length, avoided merely restating code, were easily scannable, and adopted a guiding tone. However, some preferences were mutually exclusive, such as the use of imperative mood. Stack traces were often insufficient to identify all steps developers found relevant. Developers also trusted descriptions they perceived as human-written more than those they believed were AI-generated. Finally, LLM-generated annotations of tour quality were unreliable: sycophancy, confabulation, and incoherence were pervasive.
\abstractsection{Conclusion}
This work lays a basis for future research on fine-tuning open-weight models for code tour generation, personalizing generation to accommodate diverging preferences, selecting relevant steps beyond stack traces, calibrating users' trust to avoid both disuse and misuse, and improving open-weight LLMs' ability to be more trustworthy evaluators
\end{abstract}




\begin{NoHyper}
\maketitle
\end{NoHyper}

\input{01_introduction}

\input{02_background}
\input{03_approach}

\input{04_results}

\input{05_discussion}

\input{06_conclusion}

\appendix

\input{98_appendix}

\printcredits

\section*{Declaration of generative AI and AI-assisted technologies in the manuscript preparation process. }

During the preparation of this work, the authors used AI-assisted technologies to (i) polish the language and correct the grammar and LaTeX formatting of the manuscript (Mammouth.ai and Grammarly), and (ii) assist in developing the pipeline and the web interface (Mammouth.ai and GitHub Copilot). All content generated with the assistance of these tools was thoroughly reviewed and edited by the authors to ensure it aligned with their intended meaning and contributions. The authors take full responsibility for the content of the published article.

\section*{Data availability}

The code, data, and instructions needed to replicate the experiments are available on Zenodo~\cite{replication} and GitHub (\url{https://github.com/balfroim/HumanFactorsCodeTour}).

\bibliographystyle{cas-model2-names}
\bibliography{99_references}

\bio{}
Martin Balfroid is a PhD student at the University of Namur. His research investigates AI-in-the-loop approaches to improve software engineering. He earned his master's degree in Computer Science, with a focus on Data Science, in June 2022. Martin began his PhD in July 2022 with funding from the ARIAC project and is supervised by Assistant Professors Benoît Vanderose and Xavier Devroey. 
\endbio

\bio{}
After an initial career as a librarian, Julien Albert transitioned to computer science and obtained a master's degree from UNamur in 2020. He then worked for one year at UNamur on the EFFaTA-MeM research project, which aims to develop innovative tools for text analysis. In September 2021, he began a Ph.D. in computer science at UNamur under the supervision of Professors Benoît Frenay and Bruno Dumas. His research area is explainability in artificial intelligence. His approach places the user at the center of the concerns by combining machine-learning explainability techniques with methods developed in human-computer interaction.
\endbio

\bio{}
Dzenatan Aliti is a master's student at the University of Namur. He was a research trainee at the SNAIL Team, working on software testing education through gamification.
\endbio

\bio{}
Xavier Devroey is an associate professor of software engineering at the University of Namur, where, together with Benoît Vanderose, he co-leads the SNAIL Team. His main research interests include automated software testing, test suite generation and augmentation, DevOps, software testing, and variability-intensive systems. Xavier has served as program committee member in various software engineering conferences, including ICST, ASE, EASE, MSR, and as reviewer for various international journals, including JSS, EMSE, TSE, and STVR.
\endbio

\bio{}
Benoît Vanderose is an associate professor of software engineering at the University of Namur, where he co-leads the SNAIL Team with Xavier Devroey. His main research interests include Agile software development, DevEx, and software quality.
\endbio

\end{document}

%% file: cmd/errors.tex
\newcommand{\assertJava}{Assertion}
\newcommand{\assertTest}{Assertion} 
\newcommand{\outBounds}{Out of Bounds}
\newcommand{\comparison}{Comparison}
\newcommand{\nullP}{Null Pointer}

%% file: 01_introduction.tex
\section{Introduction}
\label{sec:introduction}

Onboarding to an unfamiliar codebase is something every developer has to experience. This can be done through mentoring~\cite{ju2021case} by developers familiar with the codebase, but they can spend around a third of their time mentoring new developers~\cite{azanza2024can}. Generative AI could offload by generating interactive, context-rich documentation, such as code tours~\cite{balfroid2024towards, balfroid2025generating, kara2026lacy}, which support onboarding~\cite{taylor2022tour}. A code tour~\cite{CodeTour} consists of sequential steps, each linked to a code segment and accompanied by a summary. \Cref{fig:ct-tutorial} illustrates a tour designed to help developers understand a \texttt{NullPointerException}. (1) The first step explains the failing test \texttt{testLabelFormat}, which checks label formatting but throws a \texttt{Null\-Pointer\-Exception}. (2) The next step leads us to the \texttt{render} function, which delegates string formatting to a \texttt{Label\-Formatter}. (3) The final step identifies the faulty \texttt{format} method, which delegates label formatting to an uninitialized \texttt{Formatter}, resulting in a \texttt{NullPointerException}.

To answer this question, we conducted semi-structured interviews with 26 developers with varying levels of professional and Java experience. Each participant was confronted with six synthetically generated code tours of real Java bugs mined from GitHub commits~\cite{saavedra2024gitbugactions} from 2025. Each tour was authored by one LLM and evaluated by the two other LLMs as judges~\cite{zheng2023judging, wang2025judge} on three criteria~\cite{balfroid2024towards}: (Transparency) understanding the codebase; (Scrutability) assessing the correctness and relevance of the explanation; (Efficiency) performing the task efficiently. Participants were asked to rate each rationale on a 1--7 scale and, importantly, to think aloud~\cite{lallemand2015methodes} while doing so. Three independent coders (the first three authors) then performed qualitative coding to extract recurring experiences from the transcripts. The experiences were formulated in first person (following \citet{lejeune2019manuel}) and were further decomposed into four components: the \textit{goal} (e.g., understanding), the \textit{valence} (positive, mixed, or negative), the tour's \textit{component} it relates to (e.g., description), and the component's \textit{property} that affects the goal (e.g., the level of detail in the description).

  \input{tours/tuto}

%% file: tours/tuto.tex
\begin{codetour}{3}{The tutorial code tour. It is designed to help a developer understand a \texttt{NullPointerException}. It is an interactive document that guides the developer from the failing test case through the rendering logic to the uninitialized variable responsible for the crash. }{fig:ct-tutorial}

  \begin{codestep}{StatisticalBarRendererTest.java -- testLabelFormat()}
    \begin{lstlisting}[style=javastyle]
public class RendererTest extends TestCase {
    public void testLabelFormat() {
        StatisticalBarRenderer renderer = new StatisticalBarRenderer();
        renderer.render(); 
    }
}
    \end{lstlisting}
    \begin{stepnote}
    This step initiates the trace within \pinktt{RendererTest.java}. The \pinktt{testLabelFormat} unit test is designed to verify that the \pinktt{StatisticalBarRenderer} correctly processes label formatting. Currently, this test triggers a \pinktt{NullPointerException}. We will now examine the \pinktt{render()} method to identify the point of failure.
    \end{stepnote}
  \end{codestep}

  \begin{codestep}{StatisticalBarRenderer.java -- render()}
    \begin{lstlisting}[style=javastyle]
public class StatisticalBarRenderer {
    private LabelFormatter label = new LabelFormatter();
    public void render() {
        double mean = calculateMean();
        double std = calculateStdDev();
        String fmt = label.format("%f+-%f", mean, std);
        System.out.println(fmt);
    }
}
    \end{lstlisting}
    \begin{stepnote}
      We are now inside the \pinktt{render()} implementation of \pinktt{StatisticalBarRenderer.java}. The method calculates the mean and standard deviation before delegating string formatting to the \pinktt{LabelFormatter} instance. The execution flow encounters an issue during the call to \pinktt{label.format()}. We will step into this method to inspect the internal state of the formatter.
    \end{stepnote}
  \end{codestep}

  \begin{codestep}{StringFormatter.java -- format()}
    \begin{lstlisting}[style=javastyle]
public class LabelFormatter {
    private java.util.Formatter fmtr; 
    public String format(String fmt, Object... args) {
        return this.fmtr.format(fmt, args).toString();
    }
}
    \end{lstlisting}
    \begin{stepnote}
      The root cause is located within \pinktt{LabelFormatter.java}. The \pinktt{format} method attempts to delegate the operation to the private member \pinktt{fmtr}. However, \pinktt{fmtr} is declared but remains uninitialized. This results in a \pinktt{NullPointerException} when \pinktt{this.fmtr.format()} is invoked. To resolve this, \pinktt{fmtr} must be properly instantiated within the constructor or at the point of declaration.
    \end{stepnote}
  \end{codestep}

\end{codetour}

%% file: 02_background.tex
\section{Background}
\label{sec:background}

\begin{table}
\centering
\tiny
\caption{Positioning of this work relative to prior work on code tours. No prior study has jointly evaluated fully AI-generated code tours (\textbf{AI}) using open-weight models (\textbf{OW}), with attention to developer experience (\textbf{DX}), trust calibration (\textbf{Trust}), a debugging focus (\textbf{Debug}), and LLM-as-a-judge evaluation (\textbf{Judge}) within the pipeline. \checkmark\ indicates important coverage, $\sim$ indicates partial coverage (mentioned but not empirically evaluated), and an empty cell indicates no coverage.}
\label{tab:related-work-gap}
\begin{tabular}{rcccccc}
\toprule
\textbf{Study} & \textbf{AI} & \textbf{OW} & \textbf{DX} & \textbf{Trust}  & \textbf{Debug} &  \textbf{Judge} \\
\midrule
\citet{taylor2022tour} (2022) & & & \checkmark  & &\\
\citet{balfroid2024towards} (2024) & \checkmark & & & & \checkmark &     \\
\citet{balfroid2025generating} (2025) &  $\sim$ & $\sim$ & & & $\sim$ & $\sim$ \\
\citet{kara2026lacy} (2026) &  $\sim$ & $\sim$ & $\sim$ &  & $\sim$  &   \\
\midrule
\textbf{This work} & \checkmark & \checkmark & \checkmark & \checkmark & \checkmark & $\checkmark$  \\
\bottomrule
\end{tabular}%
\end{table}

Onboarding is the process through which new employees become active members of an organization~\cite{bauer2011organizational}. New developers usually encounter obstacles such as limited documentation, unfamiliar workflows, and new technologies~\cite{matturro2017difficulties}. Code tours are one of the many software solutions for onboarding~\cite{santos2025software}: they are structured, in-IDE walkthroughs of key code segments that guide new developers through a codebase interactively~\cite{taylor2022tour}. As a form of structured documentation, code tours directly address a lack of documentation, one of the primary onboarding obstacles as identified by \citet{matturro2017difficulties}. Beyond documentation, code tours also reduce the burden on senior developers. Indeed, mentorship is a common approach~\cite{ju2021case}, but it can consume up to 30\% of the mentor time in some large projects~\cite{schuszter2024increasing} and 10-20 hours per hire~\cite{kara2026lacy}. Rather than repeating the same walkthrough for each new hire, code tours crystallize an expert's walkthrough of the codebase into a reusable artifact~\cite{kara2026lacy}, allowing it to be authored once and reused across newcomers. \Cref{tab:related-work-gap} positions this work relative to prior work on code tours~\cite{taylor2022tour, balfroid2024towards, balfroid2025generating, kara2026lacy} and highlights the gap that no prior study has jointly evaluated fully AI-generated code tours (\textbf{AI}, \Cref{sec:bg:genai}) using open-weight models (\textbf{OW}, \Cref{sec:bg:ow}), with attention to developer experience (\textbf{DX}, \Cref{sec:bg:dx}), trust calibration (\textbf{Trust}, \Cref{sec:bg:trust}), a debugging focus (\textbf{Debug}, \Cref{sec:bg:debug}), and LLM-as-a-judge evaluation (\textbf{Judge}, \Cref{sec:bg:judge}) within the pipeline.

\subsection{AI-generated Code Tours}
\label{sec:bg:genai}

\textit{CodeTour}~\cite{CodeTour} is a VS Code extension for creating and navigating guided tours of a project within the IDE, traditionally authored by hand as a sequence of annotated code locations in JSON. Although research on code tour generation is sparse~\cite{taylor2022tour, balfroid2024towards, balfroid2025generating, kara2026lacy}, a code tour is essentially a sequence of linked code summaries, i.e., natural-language descriptions of code snippets~\cite {zhang2022survey}. Code summarization typically follows a three-step process -- modeling the source code, generating summaries, and evaluating the quality of the summaries -- and is currently dominated by machine-learning-based techniques~\cite{zhang2022survey}. Here we introduce a few notable prior works that will feed the discussion (\Cref{sec:discussion}). In a large study ($\sim$1000 participants), \citet{leinonen2023llm-vs-student} found that students rated GPT-3-generated code summaries as more understandable and accurate than student-authored ones (though the effect was small). There was no difference in length -- either objectively (number of characters) or subjectively (participant ratings) -- between student-authored and AI-authored descriptions. \citet{macneil2023experiences} studied three LLM-generated explanation types -- line-by-line, key concepts list, high-level summary -- within an interactive e-book for a web development course (58 participants). Reading time increased with code complexity, and explanations were most valued when students were unfamiliar with the code. The study was underpowered to conclude that perceived usefulness differed statistically significantly across explanation types. Looking at descriptive statistics: summary and concepts (median of 4) might be perceived as more useful than line-by-line (median of 3.5), though to an extent that is too small to be noticed with 58 participants.

Now, back to code tours, \citet{balfroid2024towards} proposed automatically generating code tours directly from stack traces. Thus, using the trace to select code locations and an LLM to author the explanations. \textit{Lacy}~\cite{kara2026lacy} extends this into a hybrid human-AI system. They shift away from a fully AI-generated pipeline by introducing a voice-to-tour feature, in which experts conduct a live walkthrough with an onboardee. Then, AI summarizes it into a reusable tour for newcomers. Although expert-curated code tours lead to measurably better learner comprehension than AI-only tours, developers still perceive fully AI-authored content as sufficient, failing to recognize what is missing. Expert-curated and AI-generated tours are nonetheless considered complementary~\cite{kara2026lacy}, making the study of AI-generated tours relevant.  Although we agree with Kara and {\.I}smail et al.~\cite{kara2026lacy} that, ideally, a code tour should be curated by an expert. In practice, this might not be feasible due to the scale. Also, there may not even be any experts involved when it comes to legacy or AI-generated code. Thus, we continue the line of work of fully AI-generated tours~\cite{balfroid2024towards, balfroid2025generating}.

\subsection{Open Weight (OW) for Code Tour Generation}
\label{sec:bg:ow}

Prior work on code tour generation~\cite{balfroid2024towards, kara2026lacy} has predominantly relied on LLMs accessed through proprietary external APIs, such as GPT-3.5 and Gemini 2.5 Flash. This reliance raises significant concerns about data privacy, as sensitive source code must be transmitted to third-party servers~\cite{azanza2024can}, and about reproducibility, as the opacity of cloud-based models impedes the ability to replicate and verify experimental results~\cite{sallou2024breaking}. A promising alternative lies in open-weight LLMs, i.e., Large Language Models whose parameters are publicly available. Thus, they can be deployed locally, thereby eliminating exposure of data to third parties and granting full control over the inference pipeline~\cite{balfroid2025generating}. Although generally smaller in scale, these models have demonstrated competitive performance against their larger, API-based counterparts, particularly in low-resource settings~\cite{wolfe2024laboratory}, suggesting that local deployment does not sacrifice much in terms of generation quality.

In this study, we will consider three open-weight models released in 2025 that specialize in software engineering: \qwCode, \dskCode, and \mistDev. There are several benchmarks to assess models' capabilities on tasks. To compare the selected models, we selected two benchmarks that together cover complementary capabilities relevant to generating debugging-focused code tours: code editing and code comprehension via input/output prediction. \textbf{AIDER}~\cite{aider2025benchmark} is designed to measure a model's ability to edit Python code across 133 small exercises that include markdown instructions, a stub Python code that specifies the functions or classes to be implemented, and unit tests. The task is to implement the provided function and class, following the instructions to pass the unit tests. The output is the candidate code to edit the benchmark. There are multiple configurations of edit formats: (i) the \textit{whole} format, where the entire file is returned; (ii) the \textit{diff} format, where each edit only provides the change (thus this is more token efficient). \textbf{CRUXEval}~\cite{gu2024cruxeval} (Code Reasoning, Understanding, and eXecution Evaluation) assesses code comprehension through 800 short Python functions, each paired with a known input–output example. It comes in two variants: input prediction, where the model must infer an input based on a given output, and output prediction, where the model must predict the output of running the function on a given input. \Cref{subsec:models} provides a comparison of the architectures and capabilities of the three models.

\subsection{Developer Experience (DX) with Code Tours}
\label{sec:bg:dx}

\citet{taylor2022tour} conducted a controlled experiment with 15 participants: an experimental group (7) received two hand-crafted code tours, while the control group (8) did not. They found that hand-crafted tours help newcomers navigate and understand a codebase. Although they reported participants' thoughts, they did not perform a systematic qualitative analysis of these excerpts. \citet{kara2026lacy} likewise did not analyze developer experiences in depth, instead observing developers to derive a set of design requirements for Lacy. In contrast, we conduct a systematic qualitative analysis of developer experience (\Cref{subsec:interview}).

\subsection{Trust Calibration}
\label{sec:bg:trust}

Trusting an automation system, per Lee \& See~\cite {lee2004trust}, means believing it will help you achieve your goal in an uncertain and vulnerable situation, which is the case when you are onboarding a new code base. However, not all systems are perfect, so trust should be calibrated to avoid \textit{misuse} (over-reliance on unreliable automation) and \textit{disuse} (rejection of capable automation)~\cite{lee2004trust}. \citet{kara2026lacy} noted that developers tended to perceive fully AI-authored content as sufficient without recognizing what is missing. This signals a risk of misuse: over-reliance on unreliable automation~\cite{lee2004trust}. In this study, we investigate more deeply which properties of the code tours components affect the trust of developers (\Cref{sec:results:trusting}) and discuss how we can better calibrate trust of developers towards fully AI-generated code tours (\Cref{sec:discussion:trust}, \Cref{sec:discussion:sycophancy}).

\subsection{Debugging for Onboarding}
\label{sec:bg:debug}

Developers reported greater efficiency when experimenting with code rather than passively reading documentation~\cite{dagenais2010moving}. Newcomers can experiment through small tasks such as fixing bugs, adding simple features, or writing unit tests.
Building on this, Balfroid et al.~\cite{balfroid2024towards} focused on generating code tours to explain stack traces. A stack trace records the sequence of nested method calls that lead to an exception, thereby identifying relevant steps for a debugging-focused code tour~\cite{balfroid2025generating}. While Lacy~\cite{kara2026lacy} has shifted from a purely debugging-oriented approach, we continue to focus on debugging because this narrows the study's scope and makes step selection straightforward using stack traces.

\subsection{LLM-as-a-judge}
\label{sec:bg:judge}

AI-generated tours have shortcomings~\cite{balfroid2024towards} that may go unnoticed by an overtrusting developer~\cite{kara2026lacy}, raising the question of how to verify the quality of code tours. Expert curation is infeasible at scale. Moreover, for legacy or AI-generated code, an expert may not even exist. Since code tour generation is an open-ended task, it is difficult to evaluate. The LLM-as-a-Judge paradigm\cite{zheng2023judging} addresses this by using LLMs to perform such evaluations. In software engineering, for instance, \citet{weyssow2024codeultrafeedback} apply this approach to assess alignment with non-functional requirements—such as readability, complexity, and style—in code tasks. While \citet{balfroid2025generating} proposed using LLM-as-a-judge to evaluate code tours specifically, they did not empirically evaluate it. To the best of our knowledge, this work is the first to investigate how developers experience LLM-generated code tour evaluations.

%% file: 03_approach.tex
\section{Evaluation Setup}
\label{sec:approach}

\begin{figure*}[t]
    \centering
    \includegraphics[width=0.95\linewidth]{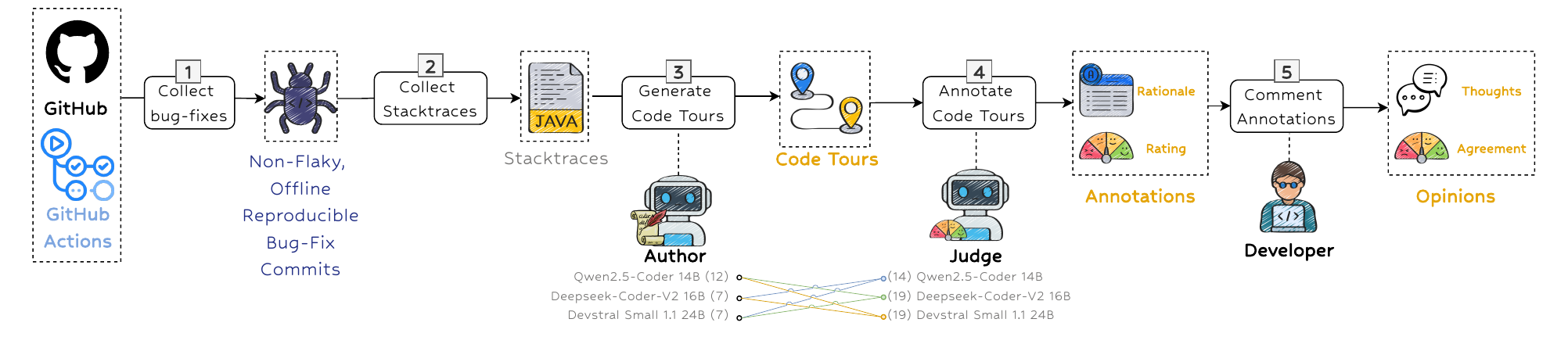}
    \caption{
    Pipeline: (1) collecting reproducible bugs using Gitbug-Actions pipeline~\cite{saavedra2024gitbugactions}, (2) executing failing tests to produce candidate stack traces, (3) generating code tour files directly from the stack trace using one of the LLMs as an author, (4) generating annotations --- a chain-of-thoughts rationale (text) and a rating (Likert-7) --- of code tours for each criterion (Transparency, Efficiency, and Scrutatibility) by each remaining LLMs as judges and (5) collecting developers opinions --- excerpts of their thoughts (text) and their agreement (Likert-7) ---  on code tours and related annotations. (Icons are from flaticon.com)}
    \label{fig:overview}
\end{figure*}

\subsection{Overview}

Now that we have reviewed the literature on code tours, it is time to present our evaluation setup for the code tour generation pipeline, the semi-structured interview, and the qualitative coding to investigate how the different properties of the components of a code tour impact the developer experience. \Cref{fig:overview} summarizes the pipeline from the bug collection to developer feedback. First (\Cref{subsec:bug_mining}), we collect reproducible bug-fix commits using the Gitbug-Actions pipeline~\cite{saavedra2024gitbugactions}. Second (\Cref{subsec:st_collect}), for each bug, we run the failing tests to collect stack traces. Third, given the stack trace and the corresponding code context, an LLM generates a structured code tour that explains the execution frames (\Cref{sec:task_gen}). Fourth, the two remaining LLMs act as judges, assigning ratings and rationales to the generated code tours based on 3 quality criteria (\Cref{sec:task_eval}). This makes 2 annotations per code tour. Finally (\Cref{subsec:interview}), 26 professional developers freely share their experiences of how the various components of a code tour (including annotations) affect them (\Cref{sec:codebook}).

\subsection{Bug Mining}
\label{subsec:bug_mining}

Balfroid et al.~\cite{balfroid2024towards} showed that it was possible to generate synthetic code tours with GPT-3.5 (closed-weight model) to explain the stack trace between a failing test and a faulty method for debugging, a common onboarding task~\cite{dagenais2010moving}. They collected stack traces from Defects4j~\cite{just2014defects4j}, a collection of 357 reproducible bugs, by instrumenting the faulty methods to record the frames at method entry, then executing the failing test to record the stack trace. However, this dataset raises concerns about data leakage~\cite{sallou2024breaking}: given LLMs' extensive training on vast internet corpora, if a source predates the training cutoff, the LLM is likely to have encountered it, biasing results. 

Gitbug-Actions~\cite{saavedra2024gitbugactions} is a tool for building up-to-date bug-fix benchmarks, specifically designed to address this problem. This team released GitBug-Java~\cite{silva2024gitbugjava}, a bug-fix benchmark of fully reproducible bugs from 2023. Nonetheless, as of 2025, this benchmark is already outdated for recent LLMs. Therefore, we analyzed 17,284 recent commits (i.e., committed in 2025) using Gitbug-Actions~\cite{saavedra2024gitbugactions} across 547 Java projects. \textbf{The mining process resulted in the collection of 110 bugs across 30 projects that can be reproduced offline and reliably across repetitions (non-flaky).}

\subsection{Stack trace Collection}
\label{subsec:st_collect}

Balfroid et al.~\cite{balfroid2024towards} used CodeQL to extract code segments corresponding to stack frames. CodeQL is a query language \textit{à la} SQL for code, performing both syntactic and semantic analysis: it builds an augmented AST and stores it in a dedicated database, which can then be queried via a proprietary DSL. The database creation step is computationally expensive and is occasionally prone to failure (the authors reported 7 failures in their study), although queries themselves are reasonably efficient. We instead use Tree-sitter to parse concrete syntax trees. This approach is significantly lighter computationally. Nevertheless, syntax trees differ in structure across languages. So, the visitor logic used to extract code segments must be reimplemented for each target language. On the other hand, CodeQL is easily portable across languages thanks to its community, which implements visitors for popular languages. We thus trade portability for efficiency.

\input{ct6-info}

We collect stack traces by running the failing tests of the 110 bugs mined using the GitBug-Java CLI~\cite{silva2024gitbugjava}. Each bug is represented as a buggy/fixed commit pair, as given by the gitbug-java info command. For instance, \Cref{fig:ct-6-info} shows the information for the commit \href{https://github.com/ezylang/EvalEx/commit/942a}{942ad41ef07c} that fixes \href{https://github.com/ezylang/EvalEx/issues/527}{Issue 527} of \href{https://github.com/ezylang/EvalEx}{EvalEx} related to a lack of handling of the evaluate function with no parameters. To identify failing tests, we use the info command to generate a list of test names that fail on the buggy commit, then parse the output to extract the individual test identifiers (\Cref{fig:ct-6-info:trace}). For each test, we run it to capture its full resulting stack trace. To determine faulty methods, we use Tree-sitter to parse the source code and extract all method definitions. We then match each method's body against the code diffs between the buggy and fixed commits, identifying any method whose body spans a diff hunk shared between them (\Cref{fig:ct-6-info:hunk}). A single bug can therefore result in multiple stack traces, for example, when several tests fail or when the fix spans across several methods. Meanwhile, some bugs produce no stack traces at all.

\textbf{From the 110 reproducible bugs from the previous step, we collected 243 stack traces across 15 projects.} Most of the errors are assertion errors (\texttt{AssertionError} or \texttt{AssertionFailedError}), and there are also \texttt{NullPointerException}, \texttt{IndexOutOfBoundsException},  and \texttt{ComparisonFailure}. The dataset is highly imbalanced, with 168 traces (69\%) originating from \texttt{jhy-jsoup} alone. To ensure fair representation across projects and error categories, we use stratified sampling per project-error types pairs: \textbf{for each evaluation round, we randomly select one stack trace per project per error type, resulting in 13 samples per round, 26 in total}.

\subsection{Models}
\label{subsec:models}

\begin{table*}[t]
    \centering
    \scriptsize

    \caption{Illustrates the architectural and capability differences among the three models.}
    \label{tab:model_specs}
    \begin{subtable}[t]{\textwidth}
        \centering

        \begin{tabular}{l l c c c c}
            \toprule
            \textbf{Model} & \textbf{Parameters} & \textbf{Knowledge Cutoff} & \textbf{Context Window} & \textbf{Size} & \textbf{Quantization} \\
            \midrule
            \href{https://ollama.com/library/qwen2.5-coder:14b/blobs/ac9bc7a69dab}{\qwCode}~\cite{hui2024qwen2.5coder} & 14B & Nov 2024~\cite{docker2025qwen} & 32K & 9 GB & Q4\_K\_M \\
            \href{https://ollama.com/library/deepseek-coder-v2:16b/blobs/5ff0abeeac1d}{\dskCode}~\cite{zhu2024deepseek} & 16B & Nov 2023~\cite{wang2025cutoff} & 160K & 8.9 GB & Q4\_0 \\
            \href{https://ollama.com/SimonPu/Devstral-Small:2507-Q4_K_XL/blobs/f352346d3b64}{\mistDev}~\cite{rastogi2025devstral} & 24B & Oct 2023~\cite{docker2025devstral} & 128K & 15 GB & Q4\_K\_M \\
            \bottomrule
        \end{tabular}
        \caption{Describe the architectural differences between the models. \textbf{Knowledge cutoff} is an estimate of the date at which the model did not train on new data. \textbf{Context Window} is the number of tokens (in thousands, K) that can be given as input. \textbf{Size} is the amount of RAM, in GB, that the models occupy. \textbf{Quantization} is the process of compressing model weights for making LLMs easier to deploy~\cite{kurt2026quantization}: \texttt{Q4\_0} is a legacy quantization format that represents each weight in 4 bits; \texttt{Q4\_K\_M} is a more efficient format that represents each weight in $\approx$ 4.5 bits with higher fidelity than \texttt{Q4\_0}. }
        \label{tab:model_specs:architecture}
    \end{subtable}
    \qquad
    \begin{subtable}[t]{\textwidth}
        \centering

        \begin{tabular}{l c c c}
            \toprule
            \textbf{Model} & \textbf{Code Editing ($\uparrow$)} & \textbf{Input Prediction ($\uparrow$)} & \textbf{Output Prediction ($\uparrow$)} \\
            \midrule
            \href{https://ollama.com/library/qwen2.5-coder:14b/blobs/ac9bc7a69dab}{\qwCode}~\cite{hui2024qwen2.5coder} & \textbf{55.7} & \textbf{72.9} & \textbf{78.8}  \\
            \href{https://ollama.com/library/deepseek-coder-v2:16b/blobs/5ff0abeeac1d}{\dskCode}~\cite{zhu2024deepseek} & 31.4 & 45.5 & 52.0   \\
            \href{https://ollama.com/SimonPu/Devstral-Small:2507-Q4_K_XL/blobs/f352346d3b64}{\mistDev}~\cite{rastogi2025devstral} & 40.7  & 64.5 & 71.5 \\
            \bottomrule
        \end{tabular}
        \caption{\textbf{Performance of selected code-generation models on code editing, execution reasoning, and real-world issue-resolution benchmarks.} \textbf{Code Editing} capabilities are assessed as the pass@1 resolve rates on the AIDER benchmark (first version~\cite{aider2025benchmark}) with the whole edit format (10 threads). \textbf{Input and Output Prediction} capabilities are evaluated on pass@1 of CRUXEval~\cite{gu2024cruxeval} with chain of thoughts (temperature set to 0.2 and 10 samples). For all benchmarks, larger numbers indicate better performance ($\uparrow$).
        }
        \label{tab:model_specs:benchmarks}
        
    \end{subtable}
    
\end{table*}

\textbf{Three open-weight agentic code models (\qwCode, \dskCode, and \mistDev) from three different providers (Alibaba, Deepseek, and Mistral), with a number of parameters ranging from 14 to 24 billion and knowledge cutoffs before 2025, were selected}. The knowledge cutoff is the date after which a model has not been trained on new data. This information is critical to consider to mitigate data leakage: if data predating the cutoff is used, models may have been trained on it, particularly for widely used datasets such as Defects4J~\cite{sallou2024breaking}. Even for open-weight models, most providers do not disclose the cutoff date for competitive reasons; community estimates are used instead. Models were accessed via the Ollama API, hosted on an on-premises cluster. The available hardware provides 48 GB of VRAM, which limits the maximum model size that can be studied. 

\Cref{tab:model_specs:architecture} provides an overview of the architecture of the three models. \qwCodeShort is the smallest model, w.r.t. the number of parameters and the context window length, but has the most recent knowledge cutoff. \dskCodeShort has the longest context window, which is the smallest in terms of size, but it uses a legacy quantization technique that produces poorer quality than the one used for the other two. \mistDevShort is the biggest in terms of size and number of parameters, but has the oldest knowledge cutoff. 

\Cref{tab:model_specs:benchmarks} provides an overview of the model's capabilities in Software Engineering, relating tasks such as code editing and input/output prediction: \qwCodeShort is the best performer, followed by \mistDevShort, while \dskCodeShort performs the poorest on those tasks.

The top-p parameter defines a threshold $p$ that picks the most probable tokens until their cumulative probability reaches at least $p$. Setting top-p to 1 ensures that all tokens are considered. A temperature of 1.0 maintains the original token probability distribution and serves as the reference setting~\cite{miller2024errorbars}. \textbf{To explore how models generate and evaluate code tour freely, we set the top-p parameter to 1 and the temperature to 1.}

\subsection{Tasks Definition}
\label{subsec:tasks}

\subsubsection{Code Tour Generation}
\label{sec:task_gen}

\begin{figure*}[t]
\centering
\begin{subfigure}{0.83\linewidth}
\centering
\fcolorbox{black}{white}{
  \begin{minipage}{0.9\linewidth}
  
  \ttfamily\footnotesize
  \textcolor{rolecolor}{%
  \promptsec{ROLE} You are a senior developer writing documentation to help onboard a new developer unfamiliar with the project domain on a debugging task.}

  \textcolor{infocolor}{%
  \promptsec{INFO} CodeTour is a Visual Studio Code extension that lets you record and play back guided walkthroughs of a codebase. 
It acts like an interactive table of contents, making it easier to: 
- onboard to a new project or feature area, 
- visualize bug reports, or 
- understand the context of a code review or pull request. 
A code tour is a sequence of interactive steps. Each step is linked to a specific directory, file, or line of code, and contains a description that explains the relevant context.}

  \textcolor{taskcolor}{%
  \promptsec{TASK} Generate a Code Tour file that walks through the execution path in the provided stack trace. Each step corresponds to a stack frame, described in order. The explanation at each step should give a clear idea of the role of the method, how it broadly works, what it produces, and explain concepts related to it. Mention potential failure causes only when they are directly relevant to the stack trace. After reading the tour, the developer should understand how the error occurs and be able to fix it. In the long term, the developer should get a better understanding of the project domain and be able to contribute more effectively.
  }

\end{minipage}
}
\caption{Generation prompt. The {\color{rolecolor}role} section sets the tone for a mentor senior developer; the {\color{infocolor}information} section provides information about code tours; and the {\color{taskcolor}task} section defines the task of generating a code tour. }
\label{fig:code-tour-gen-prompt}
\end{subfigure}
\begin{subfigure}{0.93\linewidth}
\centering

\includegraphics[width=\linewidth,trim={0mm 0mm 0mm 0mm},clip]{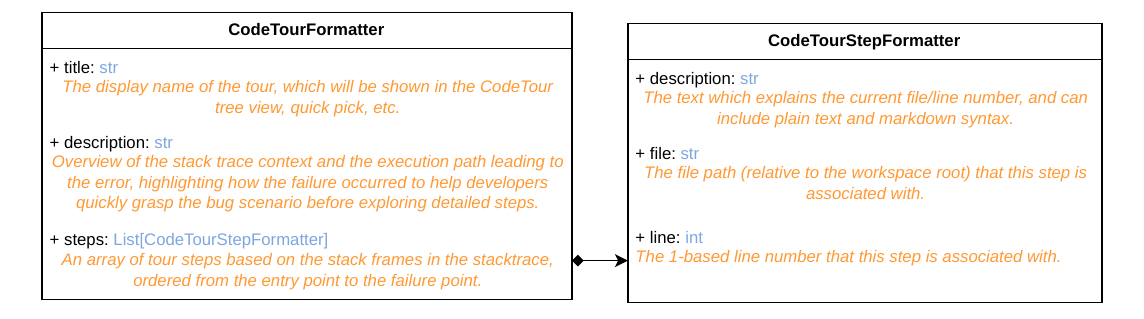}
\label{fig:code-tour-gen-schema}
\caption{Pydantic schema definition of the expected JSON structure. A \texttt{CodeTourFormatter} contains metadata and an ordered list of \texttt{CodeTourStepFormatter} entries, each pointing to a specific file and line in the codebase.}
\end{subfigure}
\caption{
Prompt components for the code tour generation task: (\subref{fig:code-tour-gen-prompt}) the prompt and (\subref{fig:code-tour-gen-schema}) the schema defining the structure in Pydantic.}
\label{fig:code-tour-gen}
\end{figure*}

Figure~\ref{fig:code-tour-gen-prompt} presents the prompt used for generating a code tour file to explain a stack trace between a faulty method and a failing test. The \textcolor{rolecolor}{\textbf{role}} section defines the persona, implicitly setting the expected completion tone~\cite{reynolds2021beyondfewshot}. The \textcolor{infocolor}{\textbf{info}} section describes the concept of a code tour, while the \textcolor{taskcolor}{\textbf{task}} section provides detailed instructions for the task.

Balfroid et al.~\cite{balfroid2024towards} generated the tour frame-by-frame, which sometimes resulted in steps lacking links. This made sense at the time because the model used in the study was GPT3.5-Turbo, which struggled with structured outputs such as JSON-formatted files~\cite{liu2024llms}. Nowadays, frameworks such as LangChain provide structured output validation. So in our approach, we generate the code tour file directly. 

\subsubsection{Code Tour Evaluation}
\label{sec:task_eval}

\Cref{fig:code-tour-eval} presents the components of the code tour evaluation task prompt. \Cref{fig:code-tour-eval-template} displays the prompt template, which can be filled in with any criterion and rating scale, and is organized into three sections. The \textcolor{rolecolor}{\textbf{role}} section defines the synthetic judge as an experienced software engineer. This serves as a cultural anchor, implicitly conveying expectations for tone and technical depth that would otherwise require lengthy, explicit instructions~\cite{reynolds2021beyondfewshot}, thereby steering the distribution of the next likely tokens. The \textcolor{taskcolor}{\textbf{task}} section states that the task is to evaluate the criterion \texttt{<CRITERIA>}, which should be filled in with the criterion's definition. \Cref{fig:checklist} lists definitions of the three criteria evaluated in this study, from~\cite{balfroid2024towards}: (Transparency) understanding the codebase; (Scrutability) assessing the correctness and relevance of the explanation; and (Efficiency) performing the task. We do not provide subcriteria to reduce potential anchoring bias. This avoids overly constraining either the model or developers with too narrow evaluation dimensions, while offering enough material for developers to respond and stimulate discussion. The definitions are slightly adjusted to be positively framed and formulated with a uniform "To what extent" structure to ensure scoring consistency. The \textcolor{formatcolor}{\textbf{format}} section defines the expected format and rating scale, using the placeholder \texttt{<RATING\_SCALE>}. \Cref{fig:rating_scale} defines the level definition used in this study: a rating of 7 indicates a very good outcome, whereas a rating of 1 indicates a poor one. The output of the task is to produce an annotation: a rating-rationale pair for a given code tour. The expected JSON schema for an annotation is defined directly in the text, not with Pydantic, since it is much simpler than the one for a code tour. It consists of two fields: "rationale" (text) and "rating" (Likert-7). We adopt a 7-point scale, as recommended by Rokeman~\cite{rokeman2024likert}, because it strikes a good balance between simplicity and granularity.

\begin{figure*}[t]
\centering

\begin{subfigure}{0.83\linewidth}
\centering
\fcolorbox{black}{white}{
  \begin{minipage}{0.9\linewidth}
  \ttfamily\footnotesize
  \textcolor{rolecolor}{\promptsec{ROLE} You are a senior software engineer reviewing a code tour.}

  \textcolor{taskcolor}{%
  \promptsec{TASK} Evaluate the quality of this code tour based on the following criterion:\\\textbf{<CRITERION>}
  }

  \textcolor{formatcolor}{%
  \promptsec{FORMAT} Provide your answer on a scale of 1~to~7.\\
  \textbf{<RATING\_SCALE>}\\
  Alongside provide a rationale for your rating.\\
  The format of your response should be in JSON:\\
\{\\
  \hspace*{1em} Rationale: "<rationale>",\\
  \hspace*{1em} Rating: "<rating>"\\
\}
}
\end{minipage}
}
\caption{Evaluation prompt template. The {\color{rolecolor}role} section sets the tone of a senior software engineer; the {\color{taskcolor}task} section defines the quality criterion to evaluate, filled with one of the definitions from~\Cref{fig:checklist}; and the {\color{formatcolor}format} section structures the output and anchors the rating scale using the levels defined in~\Cref{fig:rating_scale}.}
\label{fig:code-tour-eval-template}
\end{subfigure}

\vspace{1.5em}

\begin{subfigure}{0.83\linewidth}
\centering
\fcolorbox{black}{white}{
  \begin{minipage}{0.9\linewidth}
  \begin{footnotesize}
\textbf{1. Transparency}: To what extent does the code tour clearly explain how the code works to a new developer?
\\
\textbf{2. Scrutability}: To what extent does the code tour enable a developer to critically assess the correctness and relevance of the explanations?
\\
\textbf{3. Efficiency}: To what extent does the code tour help a developer quickly understand and navigate the codebase?
\end{footnotesize}
\end{minipage}
}
\caption{Evaluation criteria, adapted from Balfroid et al.~\cite{balfroid2024towards}. Each criterion is individually substituted into the \texttt{<CRITERION>} placeholder in the template (\Cref{fig:code-tour-eval-template}).}
\label{fig:checklist}
\end{subfigure}

\hfill 

\begin{subfigure}{0.83\linewidth}
\centering
\fcolorbox{black}{white}{
  \begin{minipage}{0.9\linewidth}
  \begin{footnotesize}
\textbf{1. [Terrible]} The code tour fails to support the criterion.\\
\textbf{2. [Very Poor]} The code tour meets the criterion only minimally.\\
\textbf{3. [Poor]} The code tour provides limited support for the criterion.\\
\textbf{4. [Basic]} The code tour meets the criterion at a minimal acceptable level.\\
\textbf{5. [Good]} The code tour supports the criterion reasonably well.\\
\textbf{6. [Very Good] }The code tour strongly supports the criterion.\\
\textbf{7. [Excellent]} The code tour fully supports the criterion.
\end{footnotesize}
\end{minipage}
}
\caption{Rating scale, substituted into the \texttt{<RATING\_SCALE>} placeholder in the template (\Cref{fig:code-tour-eval-template}).}
\label{fig:rating_scale}
\end{subfigure}
\caption{Prompt components for the code tour evaluation task: (\subref{fig:code-tour-eval-template})~the template, (\subref{fig:checklist})~the quality criteria, and (\subref{fig:rating_scale})~the rating scale.}
\label{fig:code-tour-eval}
\end{figure*}

\subsection{Developers Evaluation}
\label{subsec:interview}

Each stack trace is processed by two LLMs playing complementary roles: one acts as the author (generating a description of the code tour), and the other as the judge (rating the code tour across three criteria, with a rationale). To avoid preference leakage~\cite{li2025preference}, the author and the judge are always two different LLMs. With three LLMs, this makes six (author, judge) configurations. Thus, each sampled stack trace has two configurations, one for each of the other LLMs other than the author's LLM.

Evaluating all annotations by all developers is not feasible. To resolve this limitation, a Balanced Incomplete Block Design (BIBD)~\cite{stinson2008combinatorial} is adopted as the assignment scheme. This approach makes sure that each developer evaluates the same number $r$ of annotations, and each annotation is evaluated by the same number $k$ of developers. Additionally, every pair of developers shares the same number $\lambda$ of co-evaluated annotations. A BIBD is defined by $v$ (the number of developers), $k$ (the number of developers per annotation), and $\lambda$ (the number of co-evaluated annotations). The BIBD must satisfy two relations: $r(k-1)=\lambda(v-1)$ and $bk=vr$, where $b$ represents the total number of annotations. In this context, $b$ is the total number of rating-rationale pairs to be evaluated, $v$ is the total number of developers acting as evaluators, $r$ is the number of annotations assigned to each developer, and $k$ is the number of developers assigned to each annotation.

Pilot sessions revealed that developers require approximately 5-10 minutes to evaluate a single annotation. Given that people start to lose focus after 60 to 90 minutes, the number of annotations evaluated per developer was set to $r=6$. Accordingly, values for $k$, $\lambda$, and $v$ were determined to satisfy the BIBD relations $r(k-1)=\lambda(v-1)$ and $bk=vr$, where $b$ is the total number of annotations. The number of developers per annotation is set as $k=3$, and the number of times each pair of developers shares an annotation is set to $\lambda=1$. Consequently, the assignment scheme is of $v=13$ developers and $b=26$ annotations.

Developers were recruited based on availability, experience level, and Java proficiency. Focusing mostly on junior-level participants, as they represent the main target audience for code tours and are most likely to onboard new codebases. We replicated the assignment scheme in two parallel groups of 13 developers each. Each group was assigned a separate set of 26 annotations, structured identically according to the BIBD (See \Cref{tab:tour-authors-assigned}). Consequently, 26 developers participated, and 52 distinct annotations were evaluated. Demographics are reported in \Cref{tab:demographics}. We asked them about their highest degree in computer science: 24 have a master's, and two have a professional bachelor's. We asked them their self-reported level of development: 13 are juniors, 9 are mediors, and 4 are seniors. And also, they self-reported proficiency in Java: 12 are basic, 12 are intermediate, and 2 are advanced.

\begin{table}[t]
\centering
\scriptsize
\caption{Demographics split by round}
\label{tab:demographics}
\begin{subtable}[t]{0.48\textwidth}
\centering
\caption{Round 1}
\label{tab:round1}
\begin{tabular}{llll}
\toprule
\textbf{Participant} & \textbf{Degree} & \textbf{Dev Level} & \textbf{Java Level} \\
\midrule
P1  & Master   & Junior & Basic        \\
P2  & Bachelor & Junior & Basic        \\
P3  & Master   & Senior & Advanced     \\
P4  & Master   & Medior & Intermediate \\
P5  & Master   & Medior & Intermediate \\
P6  & Master   & Junior & Intermediate \\
P7  & Master   & Senior & Basic        \\
P8  & Master   & Junior & Intermediate \\
P9  & Master   & Medior & Intermediate \\
P10 & Master   & Junior & Basic        \\
P11 & Master   & Junior & Basic        \\
P12 & Master   & Junior & Basic        \\
P13 & Master   & Junior & Basic        \\
\bottomrule
\end{tabular}
\end{subtable}
\quad
\begin{subtable}[t]{0.48\textwidth}
\centering
\caption{Round 2}
\label{tab:round2}
\begin{tabular}{llll}
\toprule
\textbf{Participant} & \textbf{Degree} & \textbf{Dev Level} & \textbf{Java Level} \\
\midrule
P14 & Master   & Senior & Intermediate \\
P15 & Master   & Medior & Intermediate \\
P16 & Master   & Junior & Intermediate \\
P17 & Master   & Medior & Basic        \\
P18 & Master   & Medior & Advanced     \\
P19 & Master   & Junior & Basic        \\
P20 & Master   & Senior & Basic        \\
P21 & Master   & Medior & Intermediate \\
P22 & Bachelor & Junior & Intermediate \\
P23 & Master   & Medior & Intermediate \\
P24 & Master   & Medior & Intermediate \\
P25 & Master   & Junior & Basic        \\
P26 & Master   & Junior & Basic        \\
\bottomrule
\end{tabular}
\end{subtable}
\end{table}

Inside a design, author-judge configurations are distributed across developers via constraint programming: (i) the two annotations of a given stack trace must never be evaluated by the same developer, and (ii) each developer must be exposed to each configuration equally. 

The distribution of annotated generations across the three author models is unbalanced because the size of a group (13) is not divisible by the number of author models (three). Assignment is random: \qwCodeShort was assigned to 12 traces (6 in the first group,  6 in the second), \mistDevShort was assigned to 7 traces (4 in the first, 3 in the second), and \dskCodeShort was assigned to 7 traces (3 in the first, 4 in the second). \Cref{tab:tour-authors-assigned} reports which traces were assigned to which LLM as an author. Consequently, \qwCodeShort judged 14 code tours (7 from \dskCodeShort and 7 from \mistDevShort), and \dskCodeShort and \mistDevShort both judged 19 code tours (12 from \qwCodeShort and 7 from the other). Thus, as each annotation is reviewed by $k=3$ developers, we collected 42 opinions on \qwCodeShort-generated tours and 57 for the others.

\subsection{Qualitative Coding Procedure}
\label{sec:codebook}

The interviews were conducted by the first author. All interviews were conducted in French, the native language of both the participants and the interviewer, to ensure the smoothest possible communication, except for one interview conducted in English because the participant did not speak French. No audio recordings were made. 
Instead, the interviewer took notes on the fly, capturing participants' comments during the think-aloud sessions. These notes were subsequently translated into English (which we refer to as \textit{Excerpts}), and entered into a spreadsheet for analysis. Because the excerpts result from this note-taking, translation, and reformulation process, they should be understood as best-effort reconstructions of participants' reported experiences rather than verbatim transcriptions.

The qualitative coding analysis of these excerpts was done by the first three authors following an experience-based approach inspired by Lejeune et al.~\cite{lejeune2019manuel}. Each label (which we referred to as \textit{Experiences}, e.g., \nlabel{I cannot understand the code when the description lacks detail.}{E34}.) can be decomposed into four parts:
\begin{itemize}
    \item \textbf{Component} is the component of the code tour the participant refers to (e.g., the description);
    \item \textbf{Property} is the property of that component being referred to (e.g., the level of detail of the description);
    \item \textbf{Valence} is the positive or negative nature of the experience (e.g., insufficient detail is perceived negatively, whereas adequate detail is perceived positively);
    \item \textbf{Effect} is the specific aspect of the interaction with the component to which the experience relates (e.g., the description helps understand the code because of a sufficient level of detail).
\end{itemize}

After the initial transcription and labeling, the first author of this paper performed a first pass to systematize the labels. Second and third passes by the same author were conducted to systematically verify the consistency of the attributed labels. Then, independently, the second and third authors coded excerpts from their assigned participant groups (Group~1 and Group~2, respectively, each comprising 13 participants), while the first author served as the reference coder, having previously labeled the entire dataset.

We computed \kalpha\ for each \textit{Experiences}, treating the data as nominal. This coefficient corrects for expected disagreement, which is particularly appropriate in the case where there are rare \textit{Experiences}, as it is expected to agree more often because it's mostly negative for all excerpts. The three authors discussed excerpts that led to significant disagreement. After the discussion, they independently revised their codings. We repeated this cycle until all labels achieved strong agreement between authors ($\alpha \geq 0.8$).

%% file: ct6-info.tex
\begin{figure*}[t!]
\centering
\caption{Information coming from GitBug-Java for \texttt{ezylang-EvalEx-942ad41ef07c} where the first strack trace is the basis of code tour 6.}
\label{fig:ct-6-info}

\begin{subfigure}[t]{0.45\textwidth}
\begin{lstlisting}[style=trace]
### Failing Tests
- com.ezylang.evalex.functions.datetime.DateTimeFunctionsTest#testDateTimeNewNoParam
	- java.lang.AssertionError
	- 
Expecting actual throwable to be an instance of:
  com.ezylang.evalex.EvaluationException
but was:
[FAIL] DateTimeFunctionsTest#testDateTimeNewNoParam
Expected : com.ezylang.evalex.EvaluationException
Actual   : java.lang.ArrayIndexOutOfBoundsException: Index -1 out of bounds for length 0
           at DateTimeNewFunction.validatePreEvaluation(DateTimeNewFunction.java:92)
           at Expression.evaluateFunction(Expression.java:189)
           at Expression.evaluateSubtree(Expression.java:147)
\end{lstlisting}
\caption{Information about failing tests and stack traces, full stack traces are captured dynamically.}
\label{fig:ct-6-info:trace}
\end{subfigure}\hfill
\begin{subfigure}[t]{0.45\textwidth}
\begin{lstlisting}[style=diffstyle]
diff --git a/.../DateTimeNewFunction.java b/.../DateTimeNewFunction.java
--- a/.../DateTimeNewFunction.java
+++ b/.../DateTimeNewFunction.java
@@ -80,6 +80,10 @@
     int parameterLength = parameterValues.length;

+    if (parameterLength == 0) {
+      throw new EvaluationException(
+          token, "Not enough parameters for function");
+    }
+
     if (parameterLength == 1) {
       if (!parameterValues[0].isNumberValue()) {
         throw new EvaluationException(
\end{lstlisting}
\caption{Information about the patch in diff.}
\label{fig:ct-6-info:hunk}
\end{subfigure}
\end{figure*}

%% file: 04_results.tex
\section{Results}
\label{sec:results-rq1}

This section reports the results to answer the research question: "\textit{How do the properties of a code tour’s components affect the developer experience, with respect to the goals developers pursue when fixing a bug in an unfamiliar codebase?}". Following five rounds of discussion, the three coders achieved a high level of pairwise agreement ($\alpha > 0.8$) on all labels. During the sixth and final round, the coders reviewed all labels for the overall coding: they discussed possible mergers and splitters and refined the decomposition. In total, 62 labels were formulated to capture user experiences with code tours. The list of labels (denoted hereafter E1 to E62) is available in \Cref{tab:labels} (see Appendix \ref{sec:labels}). These experiences can all be clustered around five goals. Three of which are functional in nature, expressing what developers expect the code tour will help them accomplish, totaling 41 experiences (66.1\%). First, developers want to \textbf{understand} the codebase. This goal concerns 25 experiences (40.3\%), eight of which focused on \textbf{understanding \textit{efficiently}}. Second, developers want to \textbf{act} within the codebase, e.g., locating an element or fixing a bug. This goal concerns 8 experiences (12.9\%), three focusing on \textbf{acting \textit{efficiently}}). Third, developers expect to \textbf{navigate \textit{efficiently}} thanks to a code tour. This goal concerns eight experiences (12.9\%). Moreover, two goals are non-functional, relating to the properties that the developers expected of the code tour. First, developers expressed their \textbf{preferences} about code tours. Second, developers expressed what properties of the pipeline affect their \textbf{trust}. Both non-functional goals concern a total of 21 experiences (33.9\%): 10 experiences (16.1\%) for preference and 11 experiences (17.7\%) for trust. 

\begin{figure*}[t]
    \centering
    \includegraphics[width=0.9\linewidth, trim={0mm, 0mm, 0mm, 0mm}, clip]{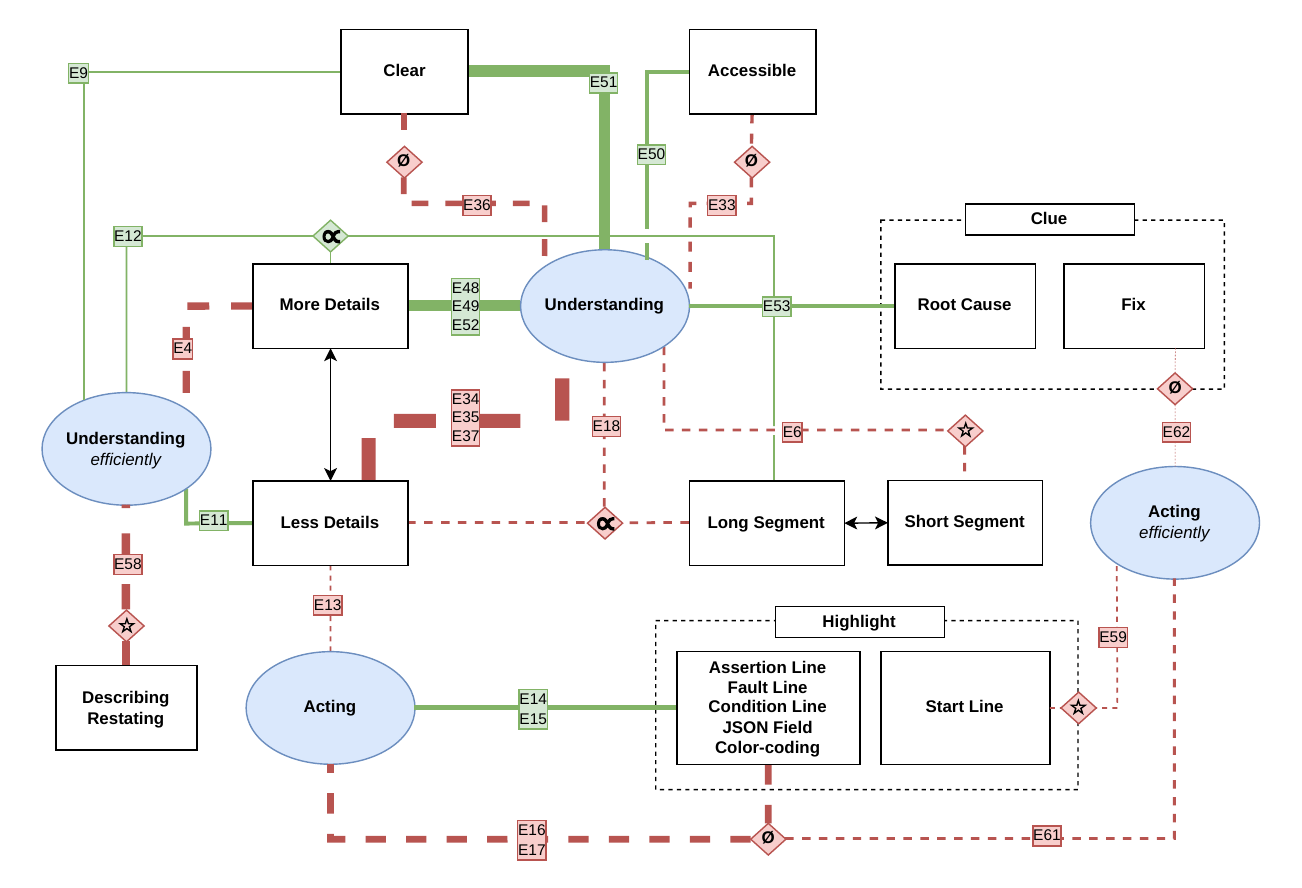}
    \caption{This diagram illustrates how the properties of the \textit{Description} component (white rectangles) impact the functional goals (blue ovals). The connecting edges represent participant experiences, marked by label IDs (Ex). Edge color indicates whether the experience was positive (green/solid) or negative (red/dashed), and edge thickness shows how many participants reported it (0.4 points per participant). The $\emptyset$ symbol shows the absence of the property. The  $\propto$  symbol shows an interaction between two properties. For instance, the E12 connection reports that for a "Long Segment", the ability to achieve "Understanding efficiently" is proportional to having "More Details." The $\medwhitestar$ relates to the value of the property, e.g., for E59, it can be understood as "referencing the start line is not needed to act efficiently."}
    \label{fig:axial-understanding-descr}
\end{figure*}

\subsection{Understanding the code base}

\subsubsection{The level of detail should be balanced between being comprehensive and concise.}
\label{sec:rq1:understanding:details}

Nearly all participants, with the exception of P21, reported \nlabel{difficulty understanding parts of the codebase at some point due to insufficient detail}{E34, E35, E37}. 23 participants cited comprehension issues with the code (E34), 9 with the background (e.g., domain concepts, acronyms, etc.) (E35), and 16 with the failure (E37). This is reflected in \Cref{fig:axial-understanding-descr} by the thick red edge linking \textit{Less Details} to \textit{Understanding}. For example, many participants, when touring the openFHIR project from medblocks, were puzzled by the acronym FHIR, which is a standard for exchanging medical data. \quotep{I don’t know what FHIR is, a small section explaining what it is would be nice.} (P11 -- 4.4). Note that four participants expressed \nlabel{a lack of detail in descriptions, but they were able to understand in the end, thanks to the judge's annotations}{E54}. Three participants indicated they would be \nlabel{unable to resolve the bug at all because the descriptions lacked sufficient detail}{E13}. Four participants noted \nlabel{the tour was missing details that the judge gave}{E28}.

Conversely, \plabel{more detailed descriptions often support participants' understanding of the codebase}{E48, E49, E52}. 17 participants noted a better understanding of the code (E48), 3 of the background (E49), and 12 of the failure (E52) as areas where the descriptions provided enough details to better understand. This is shown in \Cref{fig:axial-understanding-descr} by the thick green edge from \textit{More Details} to \textit{Understanding}. 

However, there is a balance of details to have: seven participants said that \plabel{more concise descriptions facilitated faster comprehension}{E11}, and 13 participants reported \nlabel{they are slower to understand reading a verbose description}{E4}. Nonetheless, in some cases, four participants explicitly state \mlabel{they doubt they would better understand with more or less details.}{E46}: \quotep{It's not with more examples that we would understand.} (P25 -- 2.2) and \quotep{It lacks details, but [...] I'm not sure too many details are desirable} (P8 -- 2.5).

\subsubsection{The expected level of detail scales with the code segment length.}
\label{sec:rq1:understanding:ratio}

As indicated by the $\propto$ symbol in \Cref{fig:axial-understanding-descr}, there is a proportionality between the size of the segment and the level of detail of the description. As segments became longer, three participants reported \plabel{faster comprehension when reading the summary rather than the code}{E12}, while five participants experienced \nlabel{difficulty understanding longer segments when details were insufficient}{E18}. Furthermore, five participants even found \nlabel{descriptions unnecessary for shorter segments}{E6}, noting that short methods naturally \quotep{speak for themselves} (P12 -- 1.6).

\subsubsection{Descriptions add little value when they merely restate the code.}
\label{sec:rq1:understanding:restate}

As indicated by the $\medwhitestar$ symbol in \Cref{fig:axial-understanding-descr}, a few experiences relate to properties that are not necessary. Fifteen participants reported \nlabel{being able to understand the code without reading the description}{E58}. Typically, when the author model simply restates the signature (P8 -- 4.7), lists the attributes (P6 -- 5.2), and reexplains a clear error message: there is \quotep{no need for natural language to explain an error that is clear from the beginning} (P11 -- 4.2). P7 was even more critical: \quotep{The problem is maintainability, the tour has to evolve with the code base, like every documentation, it is doomed to die after 3 weeks, it does not replace clear and maintainable code} (P7 -- G.1). P18 emphasized \quotep{I find the description superficial, it simply describes the code, reading it would be faster} (P18 -- 4.2). P26 went a bit further, and even questioned the usefulness of the overall tour for understanding: \quotep{[...] to what extent should I take into account the fact that one can understand it without going through the steps?} (P26 — 2.2).

\subsubsection{Descriptions are expected to be technical, but clear and adapted to the developer's level.}
\label{sec:rq1:understanding:clarity}

Seven participants praised certain \plabel{descriptions as accessible and junior-friendly}{E50}. Though four conceded \mlabel{their domain knowledge aided understanding}{E47}: \quotep{[...] I had an easier time understanding those pieces of code due to more familiar concepts.} (P2 -- G.3). Meanwhile, six \nlabel{struggled with descriptions not aligned with their technical level}{E33}. Difficulties most often pertained to proficiency in Java:  \quotep{The code tour was not adapted for a Java beginner.} (P1 -- G.3) or \quotep{It could have explained what a ternary operator is, [...]} (P16 -- 4.1). However, some participants also cited challenges with other technologies, for example: \quotep{Hard to understand the XML style} (P5 -- 5.1). Two participants expected technical tones when discussing code. P9 stated, \quotep{I don’t see the issue being too technical, at some point you are in code, but it is more a personal preference.} (P9 -- 2.1). P10 added, \quotep{I disagree with the transparency part, saying that the description is technical, I do not find it too technical.} (P10 -- 3.3). Moreover, P9 added about their fourth code tour \quotep{Overall, it’s a code tour. It is less fancy than the previous one, but suitable for developers, though not really for students… It’s the kind of thing I can see in a company.} (P9 -- 4.5). Twenty participants said \plabel{clear descriptions helped them understand}{E51}, as shown by the strong green edge between \textit{More Clarity} and \textit{I can understand} in \Cref{fig:axial-understanding-descr}. Three said \plabel{clarity also sped up their understanding}{E9}. Conversely, ten reported that \nlabel{unclear descriptions hindered their understanding}{E36}.

  \input{tours/18}

\subsubsection{Root cause analysis is valuable but should be given at the end.}
\label{sec:rq1:understanding:root-cause}

Seven participants \plabel{valued the inclusion of potential failure causes}{E53}, appreciating how concise root-cause summaries helped them grasp the underlying issue more efficiently: \quotep{Just reading it will help me understand root cause analysis... It's a godsend for developers... having something that sums it up in two words is great[...]} (P6 -- 2.2); \quotep{It‘s good that it explains the problem, the test, and also a potential cause of the failure … it provides a bit of guidance […] } (P8 -- 3.3); \quotep{Here it identify at least the causes, more interesting} (P15 -- 3.1); \quotep{It is good to have the potential causes […] because there was no comments in the code} (P26 -- 4.1). However, one participant strongly objected to giving a root cause analysis as it prematuraly disrupted their focus and independent reasoning, and arguing \nlabel{root cause analysis would be better placed at the end}{E20}: \quotep{[...] funny it gives potential cause in the first step … it bothers me, it gives me clue while I to understand the path [...] it makes me loose focus … it is like someone comes behind your shoulder while you are focused […] why not but at the end then, what’s more is that it is pretty vague […] it simply lost me with clues} (P18 -- 5.1).

\subsection{Acting within the code base.}

\subsubsection{Color-coding and line references (assertions, fault, condition, JSON field,...) support acting (except for referencing the start line)}
\label{sec:rq1:acting:highlight}

Nine participants reported \plabel{successfully locating an element when it was highlighted in the description}{E14}; P9 noted more specifically being \plabel{able to identify the fault because it was highlighted}{E15}. Effective cues included referencing the line number (P16 -- 6.2, among others) or specifying the field of a JSON object (P22 -- 6.3). Using color-coding, for example, highlighting methods in pink within the text (P24 -- G.2) is also appreciated (See an example in \Cref{fig:ct-32}). In contrast, seven participants were \nlabel{unable to locate an element due to insufficient highlighting}{E16}, nine \nlabel{could not identify the fault at all}{E17}, and five \nlabel{could not locate an element efficiently}{E61}. To compensate, some participants resorted to workarounds such as using Ctrl+F to find a method mentioned in the description (e.g., P13 -- 3.3). Their suggested improvements converge on a few recurring ideas: referencing line numbers (P5 -- 3.1), specifying which assertions (P8 -- 2.1) or conditions (P22 -- 4.4) failed, and visually linking the relevant line of code to the corresponding text (P11 -- 4.3). However, simply \nlabel{mentioning the segment's start line is seen as redundant}{E59} (illustrated by a $\medwhitestar$ red edge in \Cref{fig:axial-understanding-descr}). Also, much like the need for additional detail discussed earlier, the need for highlighting grows with segment length: \quotep{there is no line reference, even though the fragment is larger} (P9 -- 2.4).

\subsubsection{Suggesting a concrete fix draws mixed reactions in a debugging tour.}
\label{sec:rq1:acting:fix}

Six participants \plabel{appreciated when the tour suggested a concrete fix}{E42} and two even \plabel{trust the suggestion}{E45}: P14 found that \quotep{the solution proposition directly, it's nice [...] it seems logical as a check, but I'm not sure if you can throw an exception without changing the signature, but it is a good clue. It is clear, and it isn't a big deal if it doesn't work} (P14 -- 5.2), and P18 simply agreed that \quotep{the fix is okay, I agree with it} (P18 -- 6.2). P23 also said \nlabel{they would fix the bug faster if the description suggested a fix}{E62}

Three developers \nlabel{distrust a suggested fix that seemed incorrect}{E25}. Interestingly, in both quotes below (related to code tour in \Cref{fig:ct-9}), the suggested fix was arguably correct: the root cause was a mismatch between the assertion expecting a \texttt{UnixPath} and a method signature documented as returning a \texttt{String}, raising the legitimate question of whether the signature should be changed just to pass a test. Yet P5 went back and forth on the suggestion, stating \quotep{I disagree, your path is automatically a string... [he backtracks] actually, it's the opposite, he's right... nitpicking over the wording `supposed'... the method signature is a String... it says it's an implementation problem, but it's a documentation problem... it should say that the signature is a String} (P5 -- 4.1), while P11 (Junior, Beginner Java) similarly resisted, noting that \quotep{this method should return a path, but when the signature says it is String, you can not change it [whereas] eval was critical about this} (P11 -- 3.3). Four attendees were \nlabel{concerned about being biased by the suggestions made by the tour}{E29}

However, beyond correctness, \nlabel{some participants did not want the code tour to suggest a fix at all}{E38}, as P4 stated \quotep{I don't think the code tour should say what needs to be implemented} (P4 -- 4.1), and P25 explained that \quotep{what I like is when you do not give suggestions, you can do it yourself, while when it gives you ones, you have blinders on, you get biased} (P25 -- 6.1).

\begin{figure}[]
    \centering
    \includegraphics[width=0.9\linewidth, trim={0mm, 0mm, 0mm, 0mm}, clip]{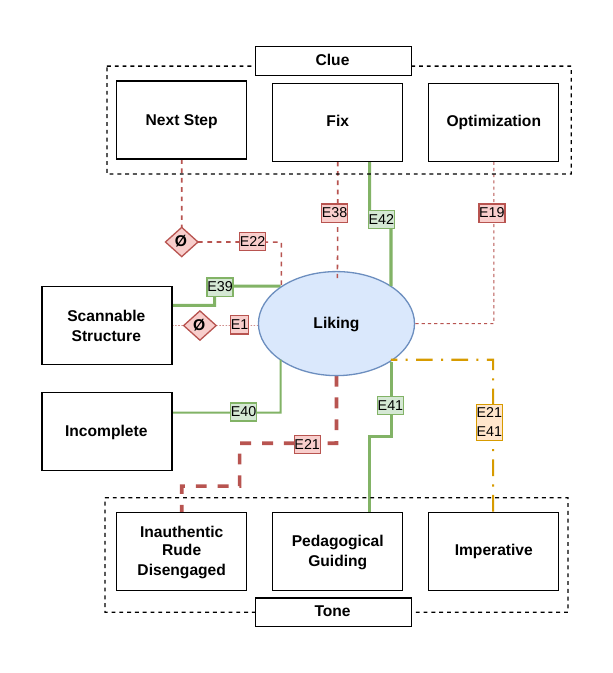}
    \caption{This diagram illustrates how the properties of the \textit{Description} component (white rectangles) impact the \textit{Liking} goals (blue ovals). The connecting edges represent participant experiences, marked by label IDs (Ex). Edge color indicates whether the experience was positive (green/solid), mixed (orange/solid-dashed), or negative (red/dashed), and edge thickness shows how many participants reported it (0.4 points per participant). The $\emptyset$ symbol shows the absence of the property.}
    \label{fig:axial-liking}
\end{figure}

\input{tours/9}

\subsection{Navigating the code base}

\subsubsection{An overview step and hinting at the next step clarify the link between steps.}
\label{sec:rq1:navigating:overview}

While seven developers reported \plabel{navigating more quickly with a well-structured tour}{E8}, six developers reported \nlabel{being slowed down in their navigation by a poorly structured tour}{E2}, ten reported being \nlabel{lost because the tour sequence was confusing}{E3}, and eight \nlabel{could not understand how the tour relates to the reported bug}{E32}, sometimes because they had to go \quotep{back and forth} between steps to understand. \quotep{[...] It's a bit difficult to get an idea of the structure [...] I don't see the link between [step] 2 and 3 [...]} (P13 -- 1.1). Moreover, while three users stated \plabel{navigating more quickly with a tour than without one}{E7}, 13 participants \nlabel{questioned the added value of the tour structure compared to tools like the IDE or terminal}{E57}. P5 remarked that \quotep{it is just like going through the stack trace, right clicking} (P5 -- 6.3), and P11 added \quotep{I'm navigating through bits of files. I have less information compared to my terminal} (P11 -- 5.3), suggesting that, in its current form, the tour does not always provide insight beyond what an IDE or terminal would already provide. 

The appropriate structure, though, would depend on the developer's goal. P15 elaborated on why for them there is \quotep{[...] two types of tour: the code tour ([...] “how this factory is initialized”) and the test tours [where] you just want to [...] go to the essential, directly to the faulty method [...] [doing] the trace in opposite direction [...] and skipped some methods that are just “waiters who serve the dishes, it is not them who added too much salt.” … the approach is different when you discover a new framework or fix a failing test. [...] code tour would be top down, test tour would be bottom up from the fault.} (P15 -- G.1). P14 also added that \quotep{[...] When I read a stacktrace, I usually go directly to the root cause by right-click, so I read the stacktrace in the other way around.} (P14 -- 1.1).

To make the structure more clear, eight participants emphasized they would \mlabel{understand faster with an overview of the tour structure}{E55}, either with a global explanation (P3 -- 3.3), a schematic explanation (P24 -- 2.2), a dependency graph (P13 -- 1.1), a navigation path (P13 -- 4.2), a mini plan (P14 -- 1.3), or a \quotep{summary of the execution tree showing the path to the error} (P15 — G.2).

Another thing that might help is highlighting the next step. Three participants expressed their \nlabel{dissatisfaction that there was no hint of where the tour is going next}{E22}: \quotep{It should indicate clearly what you will investigate next. It would be good to highlight it [because] you lose time finding the place} (P11 -- 1.1).

\subsubsection{Developers get lost more easily with tours longer than five steps.}
\label{sec:rq1:navigating:nb-steps}

Three participants reported \plabel{being faster when the tour was more concise}{E10}, while two participants reported \nlabel{being slower when the tour was longer}{E5}. P14 noted \quotep{I preferred the direct stack traces, the ones with 7 or 9 explanations, you can get lost} (P14 -- G.1), and similarly remarked \quotep{There are a lot of steps} (P14 -- 1.1) about code tour 21, which is 9 steps long. P23 further corroborated this, stating that \quotep{when you are past 4 to 5 steps, you are a bit lost} (P23 -- 1.2), while also noting that \quotep{it is less scary when you see there are three steps} (P23 -- 2.1).

\subsubsection{Duplicate steps confuse developers, yet AI judges rarely flag them.}
\label{sec:rq1:navigating:duplicates}

A few participants noticed that some tours included duplicate steps. For example, in code tour 33, P23 remarked, \quotep{the code is twice the same, but the text is different} (P23 -- 4.1), an observation echoed by P26 who described it as \quotep{the same code [...] two code the same seems like a bug} (P26 -- 5.1), and by P18 who questioned \quotep{uh the code did not change, is that normal? is it not a bug?} (P18 -- 6.1). This issue originates from the \texttt{StacktraceSkeleton} class in the tour generation pipeline (see the replication package), which does not correctly handle synthetic JVM frames. In the scenario of code tour 33, the test calls the patched method using a lambda, \texttt{assertThrows(..., () -> purl.uriDecode(...))}, which the JVM records as a synthetic frame \texttt{lambda\$invalidPercentEncoding\$N}, where \texttt{N} identifies the lambda within the method. Although developers see this as a \nlabel{redundant step that adds no value}{E57}, the LLM does not seem to view it as a critical issue.

In code tour 22, authored by \qwCodeShort, it was noted that steps 5 and 6 are duplicated, with Step 6 stating that \quotep{this step is a duplicate of the previous one. The reason being, in actual stack traces, steps could overlap, and sometimes it might cause confusion on the developer's end.}. The duplicate step also originates from a lambda call to \texttt{setField} on line 73. \dskCodeShort did not challenge the tour structure when judging the tour; no developers disagreed with the annotations, but P22 reflected: \quotep{[...] it is twice the same code [whereas] it should have been one step} (P22 -- 1.2).  However, \mistDevShort, acting as the judge, praised the duplicated steps the efficiency section \quotep{the inclusion of a duplicate step is notable but does not detract significantly from the tour's effectiveness, as it serves to emphasize a critical point in the debugging process.}. P15 disagreed, giving a 3. The judge also commended the AI-author for flagging it in terms of scrutability \quotep{the inclusion of duplicate steps is explained, which is a thoughtful addition to address potential confusion.}. P21 disagreed, giving a 2 and commented: \quotep{it's dumb, it's the same step [...]} (P21 -- 1.3).

\subsubsection{The stack trace is not enough as some interesting steps (constructors, branches, concrete implementation, or fields) are usually seen as missing steps.}
\label{sec:rq1:navigating:missing}

Seven participants reported being \nlabel{impeded in their understanding because the sequence was missing a step}{E31}. Developers expected a step about the initialization of constructors. For example, in code tour 9, the initialization of the \texttt{manager} is missing: \quotep{It lacks code to understand what is going on. For example, I can assume there is a call to a constructor somewhere upstream.} (P3 -- 5.1), though they later noted the failure is still comprehensible without it: \quotep{However, I have enough context to understand that it failed because a path and a string were compared; the code is enough, in the end.} (P3 -- 5.1). Similarly, in code tour 23, \quotep{[...] it explains well the failure, [but] the constructor is not accessible [However, it explains that] the constructor has no access [to the] protected [modifier] [...] we do not see the class in the tour [though]} (P23 -- 2.2). Some participants complained of missing steps in calls and branches: \quotep{[...] it's missing about where calls and branches are made.} (P5 -- 2.2), or a concrete implementation of some interface or abstract class: \quotep{It lacks a step when you enter the implementation of this interface} (P13 -- 6.3). Also, some tours fell short: \quotep{[...] it stops there saying the problem is there, but it does not give the numerical value [...]} (P15 -- G.3). \quotep{In the end, we know nothing about the problem, we need to go deeper, because this field has a problem, it’s just a clue, maybe it is a rabbit hole ... [the judge] said it: we do not get the whole path} (P20 -- 2.2).

\subsection{Liking}

\subsubsection{Pedagogical and guiding tones are appreciated, while disengaged, inauthentic, rude or indirect ones give rise to reservations.}
\label{sec:rq1:preference:tone}

Six developers reacted \plabel{positively to tones that felt pedagogical or guiding}{E41}. P9 valued how such framing broke from conventional documentation: \quotep{It feels like someone talking to you, not the sanitized documentation we are used to … it's a teacher, it's good for onboarding} (P9 -- 3.4).  More broadly, P9 appreciated the \quotep{variety of code tour formalisms} (P9 -- G.1), noting that \quotep{some [were] more [like] an issue, a document, a Teams conversation, or another natural conversation.} (P9 -- G.1). P10 echoed this appreciation for guidance: \quotep{I like when it says 'start' etc [...] guiding you} (P10 -- 1.1), adding that \quotep{I find it more friendly in the way it guides you step by step … Saying things like 'here the start, here the end' … simple language} (P10 -- 6.1). Such pedagogical framing, however, comes with a trade-off, as P9 observed: \quotep{[...] it is super efficient if you are a beginner, you need to be pedagogical and go into details, but for an expert, it is counterproductive.}

In contrast, participants were \nlabel{put off by tones they found disengaged, inauthentic, rude or too indirect}{E21}. Some perceived disengagement, describing the description tone as \quotep{lazy} (P6 -- 6.1) or \quotep{jaded} (P22 -- 5.9; P24 -- 4.2). They compared it to writing done out of obligation, such as when the author \quotep{was told to} (P13 -- 6.3) or {the kind of thing you would send to your boss to show it is done} (P24 -- 6.3). Others considered it inauthentic, calling it \quotep{theatrical} (P22 -- 5.4). Reading this as unreliable: \quotep{There is a tendency to oversell, it lacks sobriety, I feel like it is marketing, I would be cautious} (P13 -- 4.4). Sometimes, the description was even felt as rude: \quotep{Not very nice to speak like that. I have the impression it is someone who sent me a message … I am a bit miffed …} (P25 -- 4.1). Some participants dislike indirect phrasing, e.g., using modal verbs such as "might", as it sounds \quotep{unconfident} (P12 -- 5.2).

\subsubsection{Some developers prefer incomplete descriptions and the imperative mood, as they foster more active engagement.}
\label{sec:rq1:preference:imperative}

Interestingly, four participants \plabel{actively liked incomplete descriptions because they required more engagement with the material}{E40} (\Cref{fig:axial-liking}). As P11 noted \quotep{the lack of information allows me to understand. There is a tradeoff between too much info and not enough info because, with too much text, you do not make any effort, whereas when it gives you clues, you do.} (P11 -- 6.3). 

While \nlabel{two participants expressed discomfort with the imperative mood}{E21}, feeling they were being given orders, two others actually \plabel{received the use of the imperative mood well}{E41}. P10 said, \quotep{It is funny, it feels like it gives instructions … it is weird, why is it giving me instructions? Like orders … I don’t like it} (P10 -- 5.2, see \Cref{fig:ct-18}). P22 echoed this sentiment, noting \quotep{I feel like it gives you orders, it is peculiar, it says 'look at this constant', well give it to me then … it is its job …} (P22 -- 5.9). Others, however, read the same directness differently. P23 found it pedagogically useful despite its oddness: \quotep{It is funny that it talks to me in imperative, it’s good on a pedagogical level but weird} (P23 -- 4.2). P11 went further, identifying it as their preferred style \quotep{I don’t know if it's in the infinitive or imperative mood… Of all tours, it is the one I preferred. Here it says 'do this, does that,' so it forces you to understand the code rather than just accept it’s true. Given that I have no explanation, […], I do what I normally do, which forces me to understand.} (P11 -- 6.1, see \Cref{fig:ct-18}).

\subsubsection{Descriptions should be easily scannable using bullet points and sections.}
\label{sec:rq1:preference:text-structure}

Six participants reported \plabel{liking a text structure because it is easy to skim}{E39}. P6 liked how one tour explained \quotep{the Arrange Act Assert in bullet points} (P6 -- 4.2). P26 found a \quotep{structure in bullet points [to be] nice}, making it \quotep{clear just by reading} (P26 -- 1.3). P21 said that \quotep{list helps,} but prefers structure with subtitles, as in their third code tour (P21 -- 4.1). They added that, for the list, \quotep{it would be nice if the points were [linked to] lines of code, except, of course, if it is a one-line method. […] It would help with navigation, it helps structure your thoughts […]}. 
As for sections, P17 liked the use of subtitles (P17 -- 5.1), and P26 found \quotep{the structure is better, in 3 paragraphs} (P26 -- 4.1). \Cref{fig:ct-32} shows Code Tour 32 that illustrates a code tour with such sections.
Participants also highlighted how the structure helped them mentally process the code's logic. P12 valued the mix of \quotep{natural language and programming language,} noting that explicitly stating what a line of code does helped them \quotep{redo the logical course of the method} (P12 -- 4.2). P9 echoed this need for explicit references, stating that the structure mimicked an issue report with clear line recalls and an appreciated \quotep{quick summary of the method} (P9 -- 4.1). P12 expressed their \nlabel{dissatisfaction when a certain description structure was not easily scannable}{E1}: \quotep{[...] You would prefer a bullet point: the level of information is good but badly structured} (P12 -- 2.1).

\subsubsection{Suggestions should remain focused on the debugging task.}
\label{sec:rq1:preference:task-focus}

In Code Tour 14 (generated by \qwCodeShort), \dskCodeShort was critical about \quotep{minimal mention of optimization or efficiency gains that could be made (e.g., commenting on how certain methods like `readNumberingStyles` could potentially improve runtime performance through more efficient data structures or algorithms).}. Two participants verbally expressed their disagreement on this point: \nlabel{they do not want optimization suggestions}{E19}. P13 said \quotep{[...] we are in debugging [so] this [should] come after [because] we have to understand first. I would be lost getting discourse on optimization before understanding} (P13 -- 4.7), and rated 2 for this annotation. P4 was also critical, calling it \quotep{strange} to mention, and rated a 5, though. The third developer who had seen this annotation said nothing and even gave it a 7.

\subsection{Trusting}
\label{sec:results:trusting}

\subsubsection{Descriptions that seemed human-written were seen as trustworthy, meanwhile those seen as AI-authored were distrusted.}
\label{sec:rq1:trust:authenticity}

The more a description seemed authentic rather than synthetic, the more developers trusted its content. P9 \plabel{trust the content because it felt human-written}{E44}, therefore attributed human authorship to the text and granted it unearned authority: \quotep{the guy knows better what they are talking about than I do} (P9 -- G.1). Three admitted they \nlabel{took claims for granted without independent verification}{E60}:  \quotep{This is problematic because you take for granted what it says … you do not see what it is talking about [...]} (P11 -- 4.3); \quotep{[...] I did not check if what is written is true [...]} (P26 -- 4.2); and \quotep{Given I know nothing in XML, I suppose what it said is true … it is nice at least I understand something about style in HTML.} (P21 -- 3.2).

Conversely, right participants explicitly stated that \nlabel{believing a description was LLM-generated reduced their confidence in it}{E24}: \quotep{It makes me think of AI, so I would not trust it} (P24 -- 2.1). Usually, it is stock formulas that betray it is an LLM: \quotep{``Please let me know'' it’s busted; it is an LLM} (P12 -- 2.2). Or forcing a specific structure, such as a worthless summary final line \quotep{the last line brings nothing, LLMs sometimes reexplain the high-level point […]}. Annotations were strikingly homogeneous in phrasing and structure across the three quality criteria. Thus, after repeated exposure, participants can become harsher: \quotep{at first did not notice it was LLMs, but gradually I understood it is bullshitting me, I put more extreme values} (P18 -- G.1).

\subsubsection{Sycophancy (excessive praise), confabulations and incoherence across the pipeline impede trust.}
\label{sec:rq1:trust:issue}

13 participants pointed out \nlabel{factual inaccuracies or confabulations in the content}{E23}. Others encountered concrete errors, such as an annotation mentioning \quotep{screenshots and more detailed explanations of lines in a 2-line code snippet} (P4 -- 5.1), or emphasizing the tour for mentioning specific lines whereas \quotep{there are no line numbers contrary to what the judge said} (P23 -- 3.2).

Three participants express \nlabel{distrust when the judge contradicts itself}{E26}, e.g., \quotep{One time it says there are too many details and the other that there are not enough details} (P9 -- 5.2). P8 observed \quotep{I feel like it is not coherent … in transparency, it says it describes more or less the code but lacks details about the algorithm … but in the scrutability section, [it says] the level of details helps understand the code} (P8 -- 2.4), while P9 similarly noticed that \quotep{in the others [criteria], it says it is too technical, and here it says we understand clearly the logic behind} (P9 -- 2.3), later adding that \quotep{it seems like [it] contradicts [itself] from one criterion to the other... One time, [it] says there are too many details and the other that there are not enough details} (P9 -- 5.2). P20 further noted that \quotep{[the] comment is a bit contradictory, it says it is good and then it is not, I would have been more critical, should have said it is bad from the beginning} (P20 -- 6.1). There are also three participants who reported \nlabel{inconsistency in a tour}{E30}, for instance, \quotep{It said it delegates the task OK, then it says he won’t show here, then it shows it in the next frame} (P8 -- 2.2).

Nine participants cited \nlabel{distrust with the judge being excessively praising and positivity}{E27}: \quotep{It is obvious that it is LLM-generated because of the positive bias [and] lacks [of] negative tone, always saying it is good, even when the generator failed completely} (P5 -- G.1).

\begin{tcolorbox}[
    enhanced,
    title={How do the properties of a code tour's components affect the developer experience?},
    colback=gray!5!white,
    colframe=gray!50!black,
    fonttitle=\bfseries\sffamily,
    coltitle=white,
    colbacktitle=gray!50!black,
    boxrule=0.6pt,
    arc=2mm,
    left=6pt,
    right=6pt,
    top=6pt,
    bottom=6pt,
    breakable
]
\small

Overall, the analysis of the semi-structured interview with 26 developers shows a nuanced picture of how the properties of a code tour's components (descriptions, step sequences, judge annotations) affect the developer experience. Across 62 experiences, developers' reactions cluster around three functional goals (\textbf{understanding}, \textbf{acting}, and \textbf{navigating}) and two non-functional ones (\textbf{preference} and \textbf{trust}). 

\textbf{[Understanding]} The level of detail must be balanced (\Cref{sec:rq1:understanding:details}) and scaled to the segment length (\Cref{sec:rq1:understanding:ratio}). Descriptions should avoid merely restating the code (\Cref{sec:rq1:understanding:restate}) and are expected to be technical yet clear and accessible (\Cref{sec:rq1:understanding:clarity}). At the final step, they can provide a root cause analysis (\Cref{sec:rq1:understanding:root-cause}).

\textbf{[Acting]} Color-coding and providing line references to elements of interest (fault, failing condition, assertion, and others but not the start line) is much appreciated and helpful (\Cref{sec:rq1:acting:highlight}), while suggesting a fix is commendable, but should be done with caution due to potential bias (\Cref{sec:rq1:acting:fix}).

\textbf{[Navigating]} An overview of the tour sequence and adding hints at where the tour is going improves navigation clarity (\Cref{sec:rq1:navigating:overview}). The tour should avoid exceeding five steps (\Cref{sec:rq1:navigating:nb-steps}) and should avoid duplicate steps (\Cref{sec:rq1:navigating:duplicates}). It might also be beneficial to include steps that are not typically in the stack trace, such as constructors or the concrete implementations of an abstract class (\Cref{sec:rq1:navigating:missing}).

\textbf{[Preference]} The tone should be guiding, engaged, authentic, direct, and kind (\Cref{sec:rq1:preference:tone}). The format should be adapted to different developer personas: for example, some prefer the imperative mood or incomplete information, while others dislike it (\Cref{sec:rq1:preference:imperative}). The text should be easily scannable, using bullet points and sections (\Cref{sec:rq1:preference:text-structure}), and suggestions should stay within the task scope (\Cref{sec:rq1:preference:task-focus}). 

\textbf{[Trust]} Developers trusted descriptions that appeared human-written more than those seen as AI-authored (\Cref{sec:rq1:trust:authenticity}). Developers noted confabulatory tendencies, incoherence, and sycophantic tendencies that undermined their trust in the system (\Cref{sec:rq1:trust:issue}).

\end{tcolorbox}

%% file: tours/18.tex
\begin{codetour}{3}{\textbf{Code Tour 18 (traccar-traccar, commit \href{https://github.com/traccar/traccar/commit/de97c5099eb3}{de97c50}, trace 1/1)} generated by \qwCode. \textbf{Project.} Traccar is an open-source GPS tracking server that supports many GPS device protocols. \textbf{Commit.} This commit adds support for a new device variant (a magnetic card reader called LT32) within the H02 protocol, which devices use to send data like location to the server. \textbf{Test.} The test checks that the H02 decoder correctly handles more than 80 different message types. It fails on LT32 messages (where the binary starts with \texttt{2a} (\texttt{*}) and bytes 22–23 are \texttt{47 59} (\texttt{GY})): the decoder does not recognize them and returns nothing (\texttt{null}), which then causes a \texttt{NullPointerException} when the test tries to read the result. \textbf{Tour.} The tone of this tour is noteworthy for using the imperative mood (\textit{Understand}, \textit{Analyze}, \textit{Implement}).
}{fig:ct-18}
\begin{codestep}{H02ProtocolDecoderTest.java — testDecode()}
\begin{lstlisting}[style=javastyle, breaklines=true]
@Test
public void testDecode() throws Exception {
    // [...] Skipping two lines
    verifyAttribute(decoder, binary("2a[...20 bytes later]4759[...]"), Position.KEY_DRIVER_UNIQUE_ID, "%  ^DRIVING LICENSE$TEST$MR.^^?;6007641111111111119=180919770411=?+1419999958800100?\r\n");
    // [...] Skipping multiple lines with 83 more verify methods
}
\end{lstlisting}
\begin{stepnote}
Understand the Structure of H02 Protocol Messages
\end{stepnote}
\end{codestep}

\begin{codestep}{ProtocolTest.java — verifyAttribute()}
\begin{lstlisting}[style=javastyle, breaklines=true]
protected void verifyAttribute(BaseProtocolDecoder decoder, Object object, String key, Object expected) throws Exception {
    Object decodedObject = decoder.decode(null, null, object);
    // [...] Skipping 12 lines
}
\end{lstlisting}
\begin{stepnote}
Analyze the Decoding Process for Different Message Types
\end{stepnote}
\end{codestep}

\begin{codestep}{H02ProtocolDecoder.java — decode()}
\begin{lstlisting}[style=javastyle, breaklines=true]
@Override
protected Object decode(Channel channel, SocketAddress remoteAddress, Object msg) throws Exception {
    ByteBuf buf = (ByteBuf) msg;
    String marker = buf.toString(0, 1, StandardCharsets.US_ASCII);
    switch (marker) {
        case "*" -> // [...] Skipping 19 lines
        case "$" ->  // [...] Skipping 1 lines
        default ->  // [...] Skipping 1 lines
    }
}
\end{lstlisting}
\begin{stepnote}
Implement Attribute Verification in Tests
\end{stepnote}
\end{codestep}
\end{codetour}

%% file: tours/9.tex
\begin{codetour}{2}{\textbf{Code Tour 9 (joular-joularjx, commit \href{https://github.com/joular/joularjx/commit/9940}{9940cfc}, trace 116/150)} generated by \mistDev. \textbf{Project.} JoularX is a plugin that monitors the power consumption of Java code. \textbf{Commit.} Refactoring of the class generating the path for the results, which are now \texttt{UnixPath} instead of \texttt{String}  \textbf{Test.} The test fails because it expected a \texttt{UnixPath} and got a \texttt{String} instead. \textbf{Tour.} This is a noteworthy example of a code tour suggesting a fix.
}{fig:ct-9}
\begin{codestep}{ResultTreeManagerTest.java — getAllTotalMethodsPathTest()}
\begin{lstlisting}[style=javastyle, breaklines=true]
@Test
public void getAllTotalMethodsPathTest() {
assertEquals(Path.of(ResultTreeManager.GLOBAL_RESULT_DIRECTORY_NAME, this.appDirectory, ResultTreeManager.ALL_DIRECTORY_NAME, ResultTreeManager.TOTAL_DIRECTORY_NAME, ResultTreeManager.METHOD_DIRECTORY_NAME), this.manager.getAllTotalMethodsPath()
);}
\end{lstlisting}
\begin{stepnote}
In this step, we are setting up a test case for \pinktt{getAllTotalMethodsPath} in the \pinktt{ResultTreeManagerTest} class. The test is designed to check if the path returned by \pinktt{getAllTotalMethodsPath} matches an expected value. This method should return a \pinktt{sun.nio.fs.UnixPath} object but instead returns a \pinktt{java.lang.String}. This will be our starting point for debugging.
\end{stepnote}
\end{codestep}
\begin{codestep}{ResultTreeManager.java — getAllTotalMethodsPath()}
\begin{lstlisting}[style=javastyle, breaklines=true]
public String getAllTotalMethodsPath() { return this.allTotalMethodsPath; }
\end{lstlisting}
\begin{stepnote}
Here, we are looking at the actual implementation of \pinktt{getAllTotalMethodsPath} in the \pinktt{ResultTreeManager} class. This method is supposed to return a path to the directory where total methods consumption data is stored. However, it currently returns a string. This discrepancy will cause the assertion failure when compared with the expected value.
\end{stepnote}
\end{codestep}
\end{codetour}

%% file: 05_discussion.tex
\section{Discussion}
\label{sec:discussion}

\subsection{Implications}

\subsubsection{Aligning a model on general developers' preferences for code tour generation.}

Our findings on code tour preferences align with existing work on documentation comprehension and code foraging. This supports the intuition that a code tour is, in essence, a contextualized sequence of code summaries. Thus, code tours may be more closely related to other forms of documentation, which allows us to generalize our findings with greater confidence and draw on a broader body of prior work.

First, participants preferred balanced, high-level summaries rather than mere code restatements (\Cref{sec:rq1:understanding:details}, \ref{sec:rq1:understanding:ratio}, and \ref{sec:rq1:understanding:restate}). This corroborates prior findings that developers favor high-level overviews over line-by-line explanations~\cite{macneil2023experiences} and prefer concise comments of two to three lines~\cite{hu2022practitioners}. The ideal ratio between description and code length remains an open question for future work. It is also worth mentioning that there are techniques to reduce the presented code length rather than solely the description length, for example, by folding less-informative blocks~\cite{fowkes2017autofolding}.

Second, participants wanted scannable, structured formatting (\Cref{sec:rq1:preference:text-structure}), with color-coding and line references pointing to interesting elements (\Cref{sec:rq1:acting:highlight}), and they wanted the content to focus on the task scope (\Cref{sec:rq1:preference:task-focus}). This supports the view that, in technical documents, structure determines access to information and, consequently, affects usability. Accordingly, readers scan with the intent of answering a question~\cite{yu2026structured}, which is "how do I fix this bug ?" in the case of this study. So it is not surprising that descriptions with sections were favored, since a "Role / Execution / Potential Causes" template lets readers jump directly to the relevant part compared to a wall of text. It is also understandable that the optimization suggestions were unwelcome, as they did not answer the user's question while reading the text.

Third, developers appreciated that a guiding, engaged, authentic, direct, and kind tone (\Cref{sec:rq1:preference:tone}). Beyond the obvious fact that people prefer to be treated kindly, this echoes evidence that greater politeness within a project is associated with faster issue resolution~\cite{destefanis2016software}.

In all, these preferences appear to be broadly shared among developers. Thus, fine-tuning a model to align with them is a promising direction for future work. The findings from this work pave the way for aligning a model with developer preferences.

\subsubsection{Personalization of code tour generation}

Conversely, our findings highlight preferences that diverge across developers and might indicate distinct clusters of preferences: some prefer more agency through imperative and incomplete descriptions (\Cref{sec:rq1:preference:imperative}); developers differ in technical level (\Cref{sec:rq1:understanding:clarity}); and some are cautious about root cause analysis (\Cref{sec:rq1:understanding:root-cause}) and fix suggestions (\Cref{sec:rq1:acting:fix}). This signals a need for personalization when generating code tour descriptions.

\subsubsection{Steps selection}

This study provides a bit more insight into the kinds of steps users expect in a code tour. First, the stack trace is not enough: developers are also interested in seeing, for example, the constructor or concrete implementation of an abstract class, which are not usually present in the stack trace (\Cref{sec:rq1:navigating:missing}. Also, that lambda call related to them results in duplicate steps that developers don't want (\Cref{sec:rq1:navigating:duplicates}). However, we have to be cautious, as developers reported starting to get lost typically after five steps (\Cref{sec:rq1:navigating:nb-steps}). They also demanded an overview of the path (\Cref{sec:rq1:navigating:overview}).

\subsubsection{Trust calibration for AI-authored code tours}
\label{sec:discussion:trust}

The more a description seemed human-written rather than AI-authored, the more developers trusted its content (\Cref{sec:rq1:trust:authenticity}), which might lead to miscalibrated trust. On the one hand, there is a risk of misuse (over-reliance on unreliable automation~\cite{lee2004trust}): when some participants attributed human authorship (whereas all code tours were AI-generated), they granted it unearned authority. On the other hand, there is also a risk of disuse (rejection of capable automation~\cite{lee2004trust}): when some participants attributed AI authorship, they admitted being harsher in their judgment. 

\citet{nakano2026understanding} revealed a consistent negative shift in perceptions of trust when AI involvement in writing is disclosed, due to the loss of human touch, stylistic awkwardness, diminished author credibility, and a lack of human effort. However, assistive AI authorship can be socially acceptable when the AI contribution is under 50\%, and the writing act is descriptive. Code tour writing would fall into the \textit{Explore} category because the purpose is to develop knowledge. Another mitigating factor was the reader's AI literacy, and \citet{nakano2026understanding} recommended fostering it through a positive feedback loop by gradually exposing them to content with an increasing ratio of AI-generated parts.

Thus, trust in code tours could be calibrated by 1) disclosing what part of the descriptions were AI-generated and human-written (avoiding misuse when AI-authored content passes as human-written); 2) taking into account the AI literacy and code base and technical knowledge when assigning a tour to learners. This mitigates disuse among users with low AI literacy by gradually exposing them, and misuse among users with less knowledge who would over-trust the tour. This could be easily incorporated in onboarding systems such as Lacy~\cite{kara2026lacy}.

\subsubsection{Mitigating sycophancy, confabulation and incoherence}
\label{sec:discussion:sycophancy}

Users' trust in the code tour generation and evaluation pipeline was undermined by different issues: sycophancy (an overly positive tone), factual confabulations, and logical inconsistencies across the tour description steps and evaluation criteria (\Cref{sec:rq1:trust:issue}). You can also add to that the issue of duplicate steps discussed in \Cref{sec:rq1:navigating:duplicates} that the judge did not challenge as harshly as humans would. 

These issues overlap, e.g., participants repeatedly identified a specific confabulation in which the automated judge praised the tour for including line numbers, even though they were none. This pertains to the concept of bullshit described by \citet{hicks2024bullshit}, who argue that large language model (LLM) outputs can be characterized as bullshit because they are indifferent to truth. According to \citet{cheng2025elephant}, sycophancy refers to excessive protection of the user's ego. These concepts are interconnected: the model's disregard for truth~\cite{hicks2024bullshit} is influenced by its tendency to protect the user's image~\citet{cheng2025elephant}. Therefore, prioritizing the mitigation of sycophancy is essential to improving the trustworthiness of the pipeline.

\subsection{Threats to validity}

\subsubsection{Internal validity} 

Our excerpts are reconstructions, not verbatim transcripts. Each passes through note-taking, translation, first-person reformulation, and interpretive decisions. They should therefore be read as best-effort reconstructions of what participants reported, not as their exact words. Also, the interviews were conducted in French and translated into English, which may have introduced a loss of nuance; for instance, French has no neutral pronoun like "it" to refer to a thing, which is relevant given that the perceived human-versus-machine authorship of a description emerged as an influential signal in our study.

Because participants thought aloud while rating annotations and exploring tours, this concurrent verbalization and the presence of the interview may have heavily biased their responses. As one participant admitted \quotep{I'm not in the same mood as if I were alone [...] sometimes I answer by instinct more than by understanding} (P19 – G.1). 

Each participant evaluated six tours in a single 30- to 60-minute session without a break. So fatigue and learning effects may have influenced later ratings. Additionally, the study was conducted via a controlled web interface rather than an IDE with full access to the codebase, so behavior may differ in a realistic development setting. So, their perception of the code tour could have been very different after using it regularly for many days.

Finally, a parsing bug affecting synthetic JVM (lambda) frames produced duplicate steps that the models rarely flagged. This may have deflated agreement with the judges in those cases, as it represents an atypical scenario. However, it also constitutes a finding in its own right: it signals that the models are not critical enough of the step sequences they are given, whether acting as authors or judges.

\subsubsection{Construct validity} 

Our labels may not perfectly capture the intended theoretical concepts, due to anchoring bias during early coding, loss of nuance when forcing experiences into fixed valence categories, grouping heterogeneous excerpts under shared labels, and possible overlap between dimensions.

\subsubsection{External validity} 

Our findings may not generalize to other programming languages, to bugs committed after 2025, or to onboarding tasks beyond debugging. Recruitment primarily targeted junior developers, the population most likely to onboard to an unfamiliar codebase with external assistance and most susceptible to misuse; the number of proficient Java users in our sample is small, and more senior participants, who may be more susceptible to disuse, are underrepresented. Finally, we study agentic code LLMs with a knowledge cutoff before 2025; more recent models are likely to address some of the observed issues. However, the scarcity of reproducible bugs and data leakage as threats that must be taken seriously mean such studies inherently lag behind the state of the art.

%% file: 06_conclusion.tex
\section{Conclusion}

In this paper, we investigated how the properties of LLM-generated code tours affect developer experience when debugging unfamiliar Java codebases. We built a pipeline (\Cref{fig:overview}) that (1) mined reproducible, non-flaky bug-fix commits from 2025 GitHub repositories via Gitbug-Actions~\cite{saavedra2024gitbugactions}, (2) extracted stack traces from failing tests, (3) generates 26 code tours using three open-weight, locally hosted LLMs (\qwCode, \dskCode, and \mistDev), and (4) has the two non-authoring LLMs act as judges, producing 52 annotations on Transparency, Scrutability, and Efficiency. (5) Two groups of 13 developers evaluated each a unique set of 6 code tours and annotations while thinking aloud (each combination of tours and annotations was explored by three different developers).  Three authors qualitatively coded interviewer notes into 62 first-person experience labels (Krippendorff's $\alpha \geq 0.8$ after iterative discussion), clustered around three functional goals (understanding, acting, navigating) and two non-functional goals (liking, trusting).

We found that (1) developers want balanced detail scaled to segment length, scannable structure, guiding tone, and no mere code restatement; (2) preferences are sometimes mutually exclusive (e.g., imperative mood, deliberately incomplete descriptions); (3) the stack trace is not enough, developers would like to see snippet such as constructors; (4) perceived human authorship inflated trust (misuse risk) while perceived AI authorship deflated it (disuse risk);  and (5) LLM judges exhibited pervasive sycophancy, confabulation, and incoherence issues.

Future work should explore (1) aligning a model on general developers' preferences; (2) using personalization techniques for diverging preferences such as imperative mood; (3) studying how to select steps beyond the stack trace; (4) calibrating user trust between disuse and misuse; and (5) mitigating sycophancy, confabulation, and incoherence issues for open-weight LLMs to be trustworthier judge of code tours.

%% file: 98_appendix.tex
\input{parts/results/labels}
\input{parts/results/tour-assignements}
  \input{tours/32}


%% file: parts/results/labels.tex
\section{List of Labels}
\label{sec:labels}

\Cref{tab:labels} presents the coding guide with the different labels. Each label represents an experience from the developer's point of view and is uniquely identified (E). The table describes the developer's goal (e.g., understanding) and its valence. It also describes the component and the corresponding property it applies to (e.g., the level of detail in the description).



\begin{table*}[t]
\centering
\tiny
\caption{List of labels}
\label{tab:labels}
\begin{tabular}{lp{8cm}llllr}
\toprule
\textbf{ID} & \textbf{Name} & \textbf{Goal} & \textbf{Valence} & \textbf{Component} & \textbf{Property} & \textbf{Frequency} \\
\midrule
E1 & I dislike when the text structure is not easily scannable. & Liking & Negative & Description & Structure & 1\\
E2 & I navigate more slowly when the sequence is poorly structured. & Navigating (efficiently) & Negative & Tour & Structure & 6 \\
E3 & I navigate more slowly when the sequence is confusing. & Navigating (efficiently) & Negative & Tour & Clarity & 10 \\
E4 & I understand more slowly when descriptions are verbose. & Understanding (efficiently) & Negative & Description & Details & 13 \\
E5 & I navigate more slowly when the tour has too many steps. & Navigating (efficiently) & Negative & Tour & Length & 2 \\
E6 & I do not need a description to understand a short method. & Understanding (efficiently) & Negative & Ratio & Length & 5 \\
E7 & I navigate more quickly when the tour is well structured. & Navigating (efficiently) & Positive & Tour & Structure & 7 \\
E8 & I navigate more quickly with a tour than without one. & Navigating (efficiently) & Positive & Tour & Value & 3 \\
E9 & I understand more quickly when the descriptions are clear. & Understanding (efficiently) & Positive & Description & Clarity & 3 \\
E10 & I navigate more quickly when the tour has few steps. & Navigating (efficiently) & Positive & Tour & Length & 3 \\
E11 & I understand more quickly when the descriptions are concise. & Understanding (efficiently) & Positive & Description & Details & 7 \\
E12 & I understand long methods more quickly with a description. & Understanding (efficiently) & Positive & Description & Ratio & 3 \\
E13 & I cannot fix the bug when the descriptions lack details. & Acting & Negative & Description & Details & 3 \\
E14 & I can locate an element when it is highlighted in the description. & Acting & Positive & Description & Highlight & 9 \\
E15 & I can locate the fault when it is highlighted in the description. & Acting & Positive & Description & Highlight & 1 \\
E16 & I cannot locate an element when it is not highlighted in the description. & Acting & Negative & Description & Highlight & 7 \\
E17 & I can locate the fault when it is not highlighted in the description. & Acting & Negative & Description & Highlight & 9 \\
E18 & I cannot understand a long method when the description lacks sufficient detail. & Understanding & Negative & Description & Ratio & 5 \\
E19 & I dislike when the description gives general improvement suggestions. & Liking & Negative & Description & Suggestion & 2 \\
E20 & I dislike when the description gives root-cause clues prematurely. & Liking & Negative & Description & Timeliness & 1 \\
E21 & I dislike the description because its tone feels inauthentic, rude, or disengaged. & Liking & Negative & Description & Tone & 7 \\
E22 & I dislike when the description does not hint at the next step. & Liking & Negative & Description & Clue & 3 \\
E23 & I distrust the content because it contains factual inaccuracies or confabulations. & Trusting & Negative & Content & Accuracy & 13 \\
E24 & I distrust the content when it feels LLM-generated. & Trusting & Negative & Content & Authenticity & 8 \\
E25 & I distrust the description because it suggests a fix that appears incorrect. & Trusting & Negative & Description & Accuracy & 3 \\
E26 & I distrust the judge when it contradicts itself across criteria. & Trusting & Negative & Judge & Coherence & 3 \\
E27 & I distrust the judge when it is unconditionally positive regardless of tour quality. & Trusting & Negative & Judge & Sycophancy & 9 \\
E28 & I distrust the judge because it does not address all relevant aspects of the tour. & Trusting & Negative & Judge & Incomplete & 3 \\
E29 & I distrust the tour because its suggestions may bias my interpretation of the bug. & Trusting & Negative & Tour & Bias & 4 \\
E30 & I distrust the tour when it contradicts itself. & Trusting & Negative & Tour & Coherence & 3 \\
E31 & I cannot understand when the tour is missing a step. & Understanding & Negative & Tour & Incomplete & 7 \\
E32 & I cannot understand how the tour relates to the reported bug. & Understanding & Negative & Tour & Relevance & 8 \\
E33 & I cannot understand because the description is not adapted to my technical level. & Understanding & Negative & Description & Accessibility & 6 \\
E34 & I cannot understand the code when the description lacks details. & Understanding & Negative & Description & Details & 23\\
E35 & I cannot understand the background when the description lacks details. & Understanding & Negative & Description & Details & 9 \\
E36 & I cannot understand the description when it is unclear. & Understanding & Negative & Description & Clarity & 10 \\
E37 & I cannot understand the failure when the descriptions lack details. & Understanding & Negative & Description & Details & 16 \\
E38 & I dislike when the description suggests a concrete fix. & Liking & Negative & Description & Clue & 2 \\
E39 & I like when the text structure is easily scannable. & Liking & Positive & Description & Structure & 6 \\
E40 & I like when the description is incomplete because it forces me to engage with the code. & Liking & Positive & Description & Details & 4 \\
E41 & I like the description because its tone feels pedagogical or guiding. & Liking & Positive & Description & Tone & 6 \\
E42 & I like when the description gives general improvement suggestions. & Liking & Positive & Description & Clue & 6 \\
E43 & I understand the code more slowly because some information is given at the wrong time during the tour. & Understanding (efficiently) & Negative & Tour & Timeliness  & 3 \\
E44 & I trust the content when it feels human-written. & Trusting & Positive & Content & Authenticity & 1 \\
E45 & I trust the description because the suggested fix appears correct. & Trusting & Positive & Description & Accuracy & 2 \\
E46 & I cannot understand the code when the descriptions lack details (while the judge has them) & Understanding & Negative & Tour & Details & 4 \\
E47 & I understand because I have sufficient technical knowledge to compensate for the description. & Understanding & Mixed & Description & Accessibility & 4 \\
E48 & I understand the code when the description has enough details. & Understanding & Positive & Description & Details & 17 \\
E49 & I understand the background when the description has enough details. & Understanding & Positive & Description & Details & 3 \\
E50 & I understand the description when it is adapted to my technical level. & Understanding & Positive & Description & Accessibility & 7 \\
E51 & I understand the description when it is clear. & Understanding & Positive & Description & Clarity & 20 \\
E52 & I understand the failure when the description has enough details. & Understanding & Positive & Description & Details & 12 \\
E53 & I understand the failure because the description provides root-cause clues. & Understanding & Positive & Description & Clue & 7 \\
E54 & I doubt that I would understand better with more or less detail. & Understanding & Mixed & Description & Details & 4 \\
E55 & I would understand faster if the tour included an overview step. & Understanding (efficiently) & Negative & Tour & Overview & 8 \\
E56 & I navigate more slowly when descriptions lack details. & Navigating (efficiently) & Negative & Description & Details & 1 \\
E57 & I find some steps redundant, given what other tools (e.g., IDE or terminal) already provide. & Navigating (efficiently) & Negative & Tour & Value & 13 \\
E58 & I can understand the code without reading the descriptions. & Understanding & Negative & Description & Value & 15 \\
E59 & I do not need the descriptions to highlight an element. & Acting (efficiently) & Negative & Description & Highlight & 3 \\
E60 & I over-trust the descriptions because I assume they are correct. & Trusting & Negative & Description & Over-trust & 3 \\
E61 & I cannot locate an element quickly because it is not highlighted. & Acting (efficiently) & Negative & Description & Highlight & 5 \\
E62 & I would fix the bug faster if the description suggests a fix. & Understanding (efficiently) & Mixed & Description & Clue & 1 \\
\bottomrule
\end{tabular}
\end{table*}

%% file: parts/results/tour-assignements.tex
\section{Code tours assignments}
\label{sec:assignments}

\Cref{tab:tour-authors-assigned} presents the assignment of the generated code tours to the different participants with the experimental \textit{Group}, \textit{Tour} ID, \textit{Project}, and git commit \textit{Version}. There is a hyperlink for the tours with an example figure in the document. \textit{Author} denotes the LLM that generated the tour, and \textit{Judge} denotes the assigned LLM evaluators.
\textit{Trace} represents the stack trace index amongst all the stack traces relating to a project version, and \textit{Error} denotes the exception type. 
The \textit{Participants (Rank)} column lists the assigned human evaluators (P1--P26) and their 0-indexed evaluation order (e.g., \textit{P1[0]} means Participant 1 evaluated this tour first).

\begin{table*}[t]
\centering
\tiny
\caption{Distribution and assignment of the generated code tours.}
\label{tab:tour-authors-assigned}
\begin{tabular}{c c c c c c c c l}
\toprule
\textbf{Group} & \textbf{Tour} & \textbf{Author} & \textbf{Project} & \textbf{Version} & \textbf{Trace} & \textbf{Error} & \textbf{Judge} & \textbf{Participants (Rank)} \\
\midrule
\multirow{2}{*}{1} & \multirow{2}{*}{1} & \multirow{2}{*}{\qwCodeShort} & \multirow{2}{*}{IBM-JSONata4Java} & \multirow{2}{*}{\commit{IBM}{JSONata4Java}{e09ee334c40e}} & \multirow{2}{*}{3rd} & \multirow{2}{*}{\assertJava} 
& \dskCodeShort & P1\phantom{0} [0], P3\phantom{0} [0], P6\phantom{0} [0] \\
\lightrule
& & & & & & & \mistDevShort & P4\phantom{0} [0], P10 [0], P11 [0] \\
\midrule
\multirow{2}{*}{1} & \multirow{2}{*}{2} & \multirow{2}{*}{\mistDevShort} & \multirow{2}{*}{Nylle-JavaFixture} & \multirow{2}{*}{\commit{Nylle}{JavaFixture}{7184ab8dfacc}} & \multirow{2}{*}{1st} & \multirow{2}{*}{\assertJava} 
& \qwCodeShort & P1\phantom{0} [1], P5\phantom{0} [0], P9\phantom{0} [0] \\
\lightrule
& & & & & & & \dskCodeShort & P2\phantom{0} [0], P3\phantom{0} [1], P8\phantom{0} [0] \\
\midrule
\multirow{2}{*}{1} & \multirow{2}{*}{3} & \multirow{2}{*}{\qwCodeShort} & \multirow{2}{*}{aws-aws-lambda-snapstart-java-rules} & \multirow{2}{*}{\commit{aws}{aws-lambda-snapstart-java-rules}{beb0cf891122}} & \multirow{2}{*}{1st} & \multirow{2}{*}{\assertJava} 
& \dskCodeShort & P1\phantom{0} [2], P10 [1], P12 [0] \\
\lightrule
& & & & & & & \mistDevShort & P3\phantom{0} [2], P5\phantom{0} [1], P13 [0] \\
\midrule
\multirow{2}{*}{1} & \multirow{2}{*}{4} & \multirow{2}{*}{\dskCodeShort} & \multirow{2}{*}{classgraph-classgraph} & \multirow{2}{*}{\commit{classgraph}{classgraph}{fbe61e01e8fa}} & \multirow{2}{*}{1st} & \multirow{2}{*}{\assertTest} 
& \mistDevShort & P2\phantom{0} [1], P4\phantom{0} [1], P5\phantom{0} [2] \\
\lightrule
& & & & & & & \qwCodeShort & P3\phantom{0} [3], P9\phantom{0} [1], P10 [2] \\
\midrule
\multirow{2}{*}{1} & \multirow{2}{*}{6} & \multirow{2}{*}{\qwCodeShort} & \multirow{2}{*}{ezylang-EvalEx} & \multirow{2}{*}{\commit{ezylang}{EvalEx}{942ad41ef07c}} & \multirow{2}{*}{1st} & \multirow{2}{*}{\assertJava} 
& \dskCodeShort & P2\phantom{0} [2], P9\phantom{0} [2], P11 [1] \\
\lightrule
& & & & & & & \mistDevShort & P6\phantom{0} [1], P7\phantom{0} [0], P12 [1] \\
\midrule
\multirow{2}{*}{1} & \multirow{2}{*}{7} & \multirow{2}{*}{\dskCodeShort} & \multirow{2}{*}{fast-pack-JavaFastPFOR} & \multirow{2}{*}{\commit{fast-pack}{JavaFastPFOR}{964056681432}} & \multirow{2}{*}{1st} & \multirow{2}{*}{\assertJava} 
& \qwCodeShort & P1\phantom{0} [3], P4\phantom{0} [2], P8\phantom{0} [1] \\
\lightrule
& & & & & & & \mistDevShort & P2\phantom{0} [3], P12 [2], P13 [1] \\
\midrule
\multirow{2}{*}{1} & \multirow{2}{*}{9} & \multirow{2}{*}{\dskCodeShort} & \multirow{2}{*}{joular-joularjx} & \multirow{2}{*}{\commit{joular}{joularjx}{9940cfc00a68}} & \multirow{2}{*}{116th} & \multirow{2}{*}{\assertTest} 
& \mistDevShort & P3\phantom{0} [4], P4\phantom{0} [3], P7\phantom{0} [1] \\
\lightrule
& & & & & & & \qwCodeShort & P5\phantom{0} [3], P6\phantom{0} [2], P11 [2] \\
\midrule
\multirow{2}{*}{1} & \multirow{2}{*}{11} & \multirow{2}{*}{\qwCodeShort} & \multirow{2}{*}{medblocks-openFHIR} & \multirow{2}{*}{\commit{medblocks}{openFHIR}{fdaa43715bb5}} & \multirow{2}{*}{2nd} & \multirow{2}{*}{\comparison} 
& \dskCodeShort & P3\phantom{0} [5], P11 [3], P12 [3] \\
\lightrule
& & & & & & & \mistDevShort & P7\phantom{0} [2], P9\phantom{0} [3], P13 [2] \\
\midrule
\multirow{2}{*}{1} & \multirow{2}{*}{14} & \multirow{2}{*}{\qwCodeShort} & \multirow{2}{*}{mwilliamson-java-mammoth} & \multirow{2}{*}{\commit{mwilliamson}{java-mammoth}{b65bc063d98c}} & \multirow{2}{*}{5th} & \multirow{2}{*}{\assertTest} 
& \dskCodeShort & P4\phantom{0} [4], P6\phantom{0} [3], P13 [3] \\
\lightrule
& & & & & & & \mistDevShort & P5\phantom{0} [4], P8\phantom{0} [2], P12 [4] \\
\midrule
\multirow{2}{*}{1} & \multirow{2}{*}{15} & \multirow{2}{*}{\mistDevShort} & \multirow{2}{*}{package-url-packageurl-java} & \multirow{2}{*}{\commit{package-url}{packageurl-java}{f2e1d7aeaf7f}} & \multirow{2}{*}{9th} & \multirow{2}{*}{\assertTest} 
& \dskCodeShort & P1\phantom{0} [4], P11 [4], P13 [4] \\
\lightrule
& & & & & & & \qwCodeShort & P5\phantom{0} [5], P7\phantom{0} [3], P10 [3] \\
\midrule
\multirow{2}{*}{1} & \multirow{2}{*}{17} & \multirow{2}{*}{\mistDevShort} & \multirow{2}{*}{traccar-traccar} & \multirow{2}{*}{\commit{traccar}{traccar}{1a8487184423}} & \multirow{2}{*}{1st} & \multirow{2}{*}{\assertJava} 
& \dskCodeShort & P1\phantom{0} [5], P2\phantom{0} [4], P7\phantom{0} [4] \\
\lightrule
& & & & & & & \qwCodeShort & P6\phantom{0} [4], P8\phantom{0} [3], P9\phantom{0} [4] \\
\midrule
\multirow{2}{*}{1} & \multirow{2}{*}{\hyperref[fig:ct-18]{18}} & \multirow{2}{*}{\qwCodeShort} & \multirow{2}{*}{traccar-traccar} & \multirow{2}{*}{\commit{traccar}{traccar}{de97c5099eb3}} & \multirow{2}{*}{1st} & \multirow{2}{*}{\assertTest} 
& \mistDevShort & P2\phantom{0} [5], P6\phantom{0} [5], P10 [4] \\
\lightrule
& & & & & & & \dskCodeShort & P7\phantom{0} [5], P8\phantom{0} [4], P11 [5] \\
\midrule
\multirow{2}{*}{1} & \multirow{2}{*}{20} & \multirow{2}{*}{\mistDevShort} & \multirow{2}{*}{xxDark-jlinker} & \multirow{2}{*}{\commit{xxDark}{jlinker}{29d2f1d5a58a}} & \multirow{2}{*}{1st} & \multirow{2}{*}{\assertTest} 
& \dskCodeShort & P4\phantom{0} [5], P9\phantom{0} [5], P12 [5] \\
\lightrule
& & & & & & & \qwCodeShort & P8\phantom{0} [5], P10 [5], P13 [5] \\
\midrule
\multirow{2}{*}{2} & \multirow{2}{*}{21} & \multirow{2}{*}{\dskCodeShort} & \multirow{2}{*}{IBM-JSONata4Java} & \multirow{2}{*}{\commit{IBM}{JSONata4Java}{e09ee334c40e}} & \multirow{2}{*}{1st} & \multirow{2}{*}{\assertJava} 
& \mistDevShort & P14 [0], P16 [0], P19 [0] \\
\lightrule
& & & & & & & \qwCodeShort & P17 [0], P23 [0], P24 [0] \\
\midrule
\multirow{2}{*}{2} & \multirow{2}{*}{22} & \multirow{2}{*}{\qwCodeShort} & \multirow{2}{*}{Nylle-JavaFixture} & \multirow{2}{*}{\commit{Nylle}{JavaFixture}{7184ab8dfacc}} & \multirow{2}{*}{2nd} & \multirow{2}{*}{\assertJava} 
& \dskCodeShort & P14 [1], P18 [0], P22 [0] \\
\lightrule
& & & & & & & \mistDevShort & P15 [0], P16 [1], P21 [0] \\
\midrule
\multirow{2}{*}{2} & \multirow{2}{*}{23} & \multirow{2}{*}{\qwCodeShort} & \multirow{2}{*}{classgraph-classgraph} & \multirow{2}{*}{\commit{classgraph}{classgraph}{3ed377e7f845}} & \multirow{2}{*}{1st} & \multirow{2}{*}{\assertTest} 
& \dskCodeShort & P14 [2], P23 [1], P25 [0] \\
\lightrule
& & & & & & & \mistDevShort & P16 [2], P18 [1], P26 [0] \\
\midrule
\multirow{2}{*}{2} & \multirow{2}{*}{24} & \multirow{2}{*}{\dskCodeShort} & \multirow{2}{*}{cloudsimplus-cloudsimplus} & \multirow{2}{*}{\commit{cloudsimplus}{cloudsimplus}{0c3ecd7eb81e}} & \multirow{2}{*}{1st} & \multirow{2}{*}{\assertTest} 
& \mistDevShort & P15 [1], P17 [1], P18 [2] \\
\lightrule
& & & & & & & \qwCodeShort & P16 [3], P22 [1], P23 [2] \\
\midrule
\multirow{2}{*}{2} & \multirow{2}{*}{25} & \multirow{2}{*}{\qwCodeShort} & \multirow{2}{*}{fast-pack-JavaFastPFOR} & \multirow{2}{*}{\commit{fast-pack}{JavaFastPFOR}{964056681432}} & \multirow{2}{*}{2nd} & \multirow{2}{*}{\assertJava} 
& \dskCodeShort & P15 [2], P22 [2], P24 [1] \\
\lightrule
& & & & & & & \mistDevShort & P19 [1], P20 [0], P25 [1] \\
\midrule
\multirow{2}{*}{2} & \multirow{2}{*}{26} & \multirow{2}{*}{\qwCodeShort} & \multirow{2}{*}{jhy-jsoup} & \multirow{2}{*}{\commit{jhy}{jsoup}{78383995e7cf}} & \multirow{2}{*}{91st} & \multirow{2}{*}{\assertTest} 
& \mistDevShort & P14 [3], P17 [2], P21 [1] \\
\lightrule
& & & & & & & \dskCodeShort & P15 [3], P25 [2], P26 [1] \\
\midrule
\multirow{2}{*}{2} & \multirow{2}{*}{27} & \multirow{2}{*}{\dskCodeShort} & \multirow{2}{*}{joular-joularjx} & \multirow{2}{*}{\commit{joular}{joularjx}{9940cfc00a68}} & \multirow{2}{*}{130th} & \multirow{2}{*}{\assertTest} 
& \mistDevShort & P16 [4], P17 [3], P20 [1] \\
\lightrule
& & & & & & & \qwCodeShort & P18 [3], P19 [2], P24 [2] \\
\midrule
\multirow{2}{*}{2} & \multirow{2}{*}{29} & \multirow{2}{*}{\mistDevShort} & \multirow{2}{*}{medblocks-openFHIR} & \multirow{2}{*}{\commit{medblocks}{openFHIR}{f11766986ce4}} & \multirow{2}{*}{6th} & \multirow{2}{*}{\comparison} 
& \qwCodeShort & P16 [5], P24 [3], P25 [3] \\
\lightrule
& & & & & & & \dskCodeShort & P20 [2], P22 [3], P26 [2] \\
\midrule
\multirow{2}{*}{2} & \multirow{2}{*}{32} & \multirow{2}{*}{\qwCodeShort} & \multirow{2}{*}{mwilliamson-java-mammoth} & \multirow{2}{*}{\commit{mwilliamson}{java-mammoth}{b65bc063d98c}} & \multirow{2}{*}{4th} & \multirow{2}{*}{\assertTest} 
& \dskCodeShort & P17 [4], P19 [3], P26 [3] \\
\lightrule
& & & & & & & \mistDevShort & P18 [4], P21 [2], P25 [4] \\
\midrule
\multirow{2}{*}{2} & \multirow{2}{*}{33} & \multirow{2}{*}{\mistDevShort} & \multirow{2}{*}{package-url-packageurl-java} & \multirow{2}{*}{\commit{package-url}{packageurl-java}{f2e1d7aeaf7f}} & \multirow{2}{*}{8th} & \multirow{2}{*}{\assertTest} 
& \dskCodeShort & P14 [4], P24 [4], P26 [4] \\
\lightrule
& & & & & & & \qwCodeShort & P18 [5], P20 [3], P23 [3] \\
\midrule
\multirow{2}{*}{2} & \multirow{2}{*}{34} & \multirow{2}{*}{\mistDevShort} & \multirow{2}{*}{thibaultmeyer-cuid-java} & \multirow{2}{*}{\commit{thibaultmeyer}{cuid-java}{d923856da2fb}} & \multirow{2}{*}{1st} & \multirow{2}{*}{\assertTest} 
& \dskCodeShort & P14 [5], P15 [4], P20 [4] \\
\lightrule
& & & & & & & \qwCodeShort & P19 [4], P21 [3], P22 [4] \\
\midrule
\multirow{2}{*}{2} & \multirow{2}{*}{35} & \multirow{2}{*}{\qwCodeShort} & \multirow{2}{*}{traccar-traccar} & \multirow{2}{*}{\commit{traccar}{traccar}{a16ad4de29b3}} & \multirow{2}{*}{1st} & \multirow{2}{*}{\outBounds} 
& \mistDevShort & P15 [5], P19 [5], P23 [4] \\
\lightrule
& & & & & & & \dskCodeShort & P20 [5], P21 [4], P24 [5] \\
\midrule
\multirow{2}{*}{2} & \multirow{2}{*}{36} & \multirow{2}{*}{\dskCodeShort} & \multirow{2}{*}{traccar-traccar} & \multirow{2}{*}{\commit{traccar}{traccar}{52cab39d01d3}} & \multirow{2}{*}{1st} & \multirow{2}{*}{\nullP} 
& \qwCodeShort & P17 [5], P22 [5], P25 [5] \\
\lightrule
& & & & & & & \mistDevShort & P21 [5], P23 [5], P26 [5] \\
\bottomrule
\end{tabular}
\end{table*}

%% file: tours/32.tex
\begin{codetour}{3}{\textbf{Code Tour 32 (mwilliamson-java-mammoth, commit b65b...)} generated by \qwCode. The tour covers the parsing order of XML styles (supposedly for a Word document, since Mammoth is a Word-to-HTML converter). The test fails because of an \texttt{AssertionFailedError}, expecting \texttt{<Optional[Normal Table]>} but receiving \texttt{<Optional[Table Normal]>}. The structure of the descriptions is noteworthy because it is separated into three sections: role, execution, and potential causes.}{fig:ct-32}
\begin{codestep}{StylesXmlTests.java — whenMultipleNonNumberingStyleElementsHaveSameStyleIdThenOnlyFirstElementIsUsed()}
 \begin{lstlisting}[style=javastyle]
@Test
public void whenMultipleNonNumberingStyleElementsHaveSameStyleIdThenOnlyFirstElementIsUsed() {
    XmlElement element = element("w:styles", list(
        element("w:style", map("w:type", "table", "w:styleId", "TableNormal"), list(nameElement("Normal Table"))),
        element("w:style", map("w:type", "table", "w:styleId", "TableNormal"), list(nameElement("Table Normal")))
    ));
    Styles styles = readStylesXmlElement(element);
    assertEquals(Optional.of("Normal Table"), styles.findTableStyleById("TableNormal").get().getName());
}
    \end{lstlisting}
    \begin{stepnote}
Start by examining the failing test method in the provided Java codebase.

\textbf{Role:} The \pinktt{whenMultipleNonNumberingStyleElementsHaveSameStyleIdThenOnlyFirstElementIsUsed} method tests the behavior of the \pinktt{readStyles} function when multiple non-numbering style elements have the same style ID. Specifically, it ensures that only the first element with a given ID is used.

\textbf{Execution:} The test sets up an XML structure containing two \pinktt{w:style} elements with the same \pinktt{w:styleId} but different names (\pinktt{Normal Table} and \pinktt{Table Normal}). It then calls \pinktt{readStylesXmlElement} to process this XML and checks that the style returned by \pinktt{findTableStyleById} has a name of \pinktt{Normal Table}, which is the first encountered.

\textbf{Potential Causes:} \\
1. The \pinktt{readStyles} function might not be correctly filtering elements based on their type or ID. \\ 2. The order in which styles are processed could lead to unexpected results, such as picking the second element instead of the first.
    \end{stepnote}
\end{codestep}
  \begin{codestep}{StylesXml.java — readStylesXmlElement()}
    \begin{lstlisting}[style=javastyle]
public class StylesXml {
    public static Styles readStylesXmlElement(XmlElement element) {
        XmlElementList styleElements = element.findChildren("w:style");
        return new Styles(readStyles(styleElements, "paragraph"), readStyles(styleElements, "character"), readStyles(styleElements, "table"), readNumberingStyles(styleElements)
        );
}
    \end{lstlisting}
    \begin{stepnote}
Navigate to the \pinktt{readStylesXmlElement} method in the provided Java codebase.

\textbf{Role:} The \pinktt{readStylesXmlElement} method is a factory function that creates a \pinktt{Styles} object from an XML element. It processes different types of styles (paragraph, character, table, and numbering) by calling separate helper methods for each type.

\textbf{Execution:} The \pinktt{readStylesXmlElement} method finds all child elements with the tag \pinktt{w:style} within the provided XML element. It then invokes \pinktt{readStyles}, \pinktt{readNumberingStyles}, and other helpers to extract and store styles of different types in a \pinktt{Styles} object.

\textbf{Potential Causes:} If any style fails to be processed or an unexpected error occurs during the processing, it could potentially affect the final \pinktt{Styles} object returned by this method.

    \end{stepnote}
  \end{codestep}
  \begin{codestep}{StylesXml.java — readStyles()}
  \begin{lstlisting}[style=javastyle]
private static Map<String, Style> readStyles(XmlElementList styleElements, String styleType) {
        return toMap(styleElementsOfType(styleElements, styleType), StylesXml::readStyle);
}
  \end{lstlisting}
  \end{codestep}
   \begin{stepnote}
Navigate to the \pinktt{readStyles} method in the provided Java codebase.

\textbf{Role:} The \pinktt{readStyles} method is responsible for filtering and processing \pinktt{w:style} elements of a specific type (e.g., paragraph, character, table). It extracts and maps these styles into a \pinktt{Map<String, Style>} where each key is the style ID and each value is the corresponding Style object.

\textbf{Execution:} The \pinktt{readStyles} method takes an XML element list containing multiple \pinktt{w:style} elements and filters them based on the given \pinktt{styleType}. It then maps these filtered elements into a map of styles, using their IDs as keys by calling the \pinktt{readStyle} helper method for each element.

\textbf{Potential Causes:}

1. If there are duplicate style IDs within the filtered list, it could lead to unexpected behavior, such as overwriting existing entries in the map with new ones. This could result in a scenario where the last encountered style with a given ID is used instead of the first one.

   \end{stepnote}
\end{codetour}